\documentclass[referee]{aa} 
\usepackage{graphicx}
\usepackage{txfonts}
\usepackage{longtable}
\usepackage{lscape}
\usepackage{natbib}
\usepackage{color}
\bibpunct{(}{)}{;}{a}{}{,} 
\def\Msun{\hbox{$\thinspace M_{\odot}$}}
\def\Rsun{\hbox{$\thinspace R_{\odot}$}}
\def\Teff{\hbox{$\thinspace T_{\mathrm{eff}}$}}
\def\MJ{\hbox{$\thinspace M_{\mathrm{J}}$}}
\def\kms{\hbox{$\thinspace {\mathrm{km~s^{-1}}}$}}
\def\ms{\hbox{$\thinspace {\mathrm{m~s^{-1}}}$}}
\def\vt{\hbox{$\thinspace v_{\mathrm{t}}$}}

\begin{document}
   \title{The Penn State - Toru\'n Centre for Astronomy Planet Search stars
    \thanks{Based on observations obtained with the Hobby-Eberly Telescope, 
    which is a joint project of the University of Texas at Austin, the 		
    Pennsylvania State University, Stanford University, 
    Ludwig-Maximilians-Universit\"at M\"unchen, and Georg-August-Universit\"at 
    G\"ottingen.}
	}

   \subtitle{III. The evolved stars sample}
   
   \titlerunning{PTPS stars. III. The evolved stars sample.}

   \author{A. Niedzielski
          \inst{1}
          \and
	   B. Deka-Szymankiewicz
          \inst{1}
                    \and
	   M. Adamczyk
          \inst{1}
         \and
	   M. Adam\'ow
          \inst{2,1}
          \and
	  G. Nowak
          \inst{3,4,1}
          \and          
          A. Wolszczan
	  \inst{5,6} 
          }

   \institute{Toru\'n Centre for Astronomy, Nicolaus Copernicus University 
   in Toru\'n, Grudziadzka 5, 87-100 Toru\'n, Poland\\
              \email{Andrzej.Niedzielski@umk.pl}
            \and 
            McDonald Observatory and Department of Astronomy, University of Texas at Austin, 
            2515 Speedway, Stop C1402, Austin, Texas, 78712-1206, USA
            \and
            Instituto de Astrof\'{\i}sica de Canarias, C/ V\'{\i}a L\'actea, s/n, E38205 - La Laguna,Tenerife, Spain
            \and
            Departamento de Astrof\'{\i}sica, Universidad de La Laguna, E-38206 La Laguna, Tenerife, Spain
          \and
             Department of Astronomy and Astrophysics, 
             Pennsylvania State University, 525 Davey Laboratory, 
             University Park, PA 16802
	 \and
             Center for Exoplanets and Habitable Worlds, 
             Pennsylvania State University, 525 Davey Laboratory, 
             University Park, PA 16802           
             }

   \date{Received ; accepted }

 
  \abstract
{}
{We  intend to present complete spectroscopic analysis of 455 stars 
observed within the Penn State - Toru\'n Centre for Astronomy 
 Planet Search (PTPS) with the High Resolution Spectrograph of the 9.2~m Hobby-Eberly
 Telescope. 
 We will also present the total 
 sample of 744 evolved stars of PTPS and discuss 
 masses of stellar hosts in our and other surveys devoted to evolved planetary systems.}
{Stellar atmospheric parameters  were determined through a strictly 
spectroscopic LTE analysis of equivalent widths of Fe\,I and Fe\,II lines. 
Rotational velocities were obtained from synthetic spectra fitting.
Radial velocities were obtained from  Gaussian function fitting to the 
cross-correlation function. 
We determined stellar masses, ages  and luminosities  via Bayesian analysis of 
theoretical isochrones. The radii were calculated either from  derived masses and 
$\log g$ or from $\Teff$ and luminosities. }
{We present basic atmospheric parameters ($\Teff$, $\log g$, $\vt$
 and [Fe/H]), rotation velocities and absolute radial velocities as well as 
 luminosities, masses, ages and radii for  402 stars (including 11 single-lined spectroscopic 
binaries), mostly subgiants 
 and giants.  For 272 of them we present parameters for the first time. For another 53 stars   
 we present estimates of $\Teff$ 
 and $\log g$  based on photometric calibrations.
More than half objects were found to be subgiants, there is also a large group 
of giants  and a few stars appeard to be dwarfs. The results show that the presented sample 
is composed of stars with masses ranging from 0.52 to $3.21\Msun$ of which 17
have masses $\geq$ $2.0\Msun$. The radii of stars studied in this paper range from 0.66 to 
$36.04\Rsun$ with vast majority having radii between 2.0 and $4.0\Rsun$. They are generally less metal abundant than the Sun with median 
[Fe/H]$=-0.07$. For 62 stars in common with other planet searches we found a very good agreement in obtained stellar atmospheric parameters.
We also present basic properties of the complete list of 
744 stars that form the PTPS evolved stars sample.  
We examined stellar masses for 1255 stars in five other  planet searches and found some of them likely to be significantly overestimated.
Applying our uniformly determined stellar masses we confirm the apparent increase of companions masses for evolved stars, and we explain it, as well as lack of close-in planets  with limited effective radial velocity precision for  those stars  due to activity.
}

{}


   \maketitle
%

\section{Introduction}

After over 20 years of research, with $\sim$2000 planets found around 
other stars since discoveries of the first extrasolar systems by \citet{WF1992,MQ1995} and 
\citet{MB1996},  
it appears clear that of all available observational 
techniques applied to searches for exoplanets the precise radial velocity (RV) and stellar transits delivered most of data.

The RV technique has been proved to be especially useful in search for planets around massive and evolved stars.
Massive and intermediate-mass main-sequence (MS) stars have high effective temperatures and rotate
rapidly. Due to paucity of spectral lines and their width these  stars  are  not  suitable
for high precision RV searches for planetary  companions. 
Unfortunately, planetary candidates around such stars, even if 
discovered occasionally in transit
searches (cf. \citealt{Borucki2011, Schwamb2013}), are very difficult to confirm with
RV  measurements.
Consequently, transit projects have delivered very few planetary systems 
around stars much more massive than the Sun, with Kepler-432 \citep{Ciceri2015,Ortiz2015,Quinn2015},
Kepler-435  \citep{Almenara2015}
being the most prominent examples so far.

RV  searches
that  focus on giant and subgiant stars which are evolving off the MS, cooling down, and
considerably  slowing  their  rotation, exhibiting abundant narrow-line line spectrum 
that makes them accessible to RV technique, 
deliver most of  data  on  such  planetary  systems.   
The slowly growing population of currently known $\sim60$ 
planets around evolved stars 
is a result of intense research 
in projects like  McDonald 
Observatory Planet Search \citep{1993ASPC...36..267C, 1993ApJ...413..339H}, 
Okayama Planet Search \citep{2003ApJ...597L.157S}, 
Tautenberg Planet Search \citep{2005A&A...437..743H},
Lick K-giant Survey \citep{2002ApJ...576..478F},  
ESO FEROS planet search \citep{2003A&A...398L..19S, 2003A&A...397.1151S}, 
Retired A Stars and Their Companions \citep{Johnson2007},
Coralie $\&$ HARPS search \citep{2007A&A...472..657L},
Boyunsen Planet Search \citep{2011A&A...529A.134L}, our own 
Pennsylvania-Toru\'n Planet Search  (PTPS, \citealt{Nie2007, NW2008}),
and several others. 

The  most  massive  hosts  of  planetary  systems  come
almost exclusively from such surveys (e.g.  \citealt{Sato2007, Sato2010, Sato2012, Sato2013b}). 
They have  demonstrated, for example, a paucity of planets within 0.5 AU
of their parent stars \citep{Johnson2007, Sato2008, Jones2014}, a borderline currently broken by
Kepler~91~b \citep{Lillo-Box2014, Barclay2015}. They also proved capable of delivering evidence for recent violent 
star-planet interactions in aging planetary systems \citep{Adamow2012}.
We note, however, a slowly growing statistics of planetary systems around MS stars 
more massive than the Sun discovered in transit surveys (KELT-7  - \citealt{Bieryla2015}, 
WASP-78 - \citealt{Smalley2012}, 
HET-P-40 - \citealt{Hartman2012}). 

Stellar masses are essential in determining planetary mass companions minimum 
masses ($m_{\rm{P}} \sin i$) in all RV planet searches. Precise determinations 
of masses of isolated single MS stars are already not easy, but in the case of evolved stars, like 
subgiants or giants, they are even more difficult.  
From all indirect methods to estimate stellar mass, asteroseismology is probably the most reliable one. 
Unfortunately vast majority of  stars searched for planets with the RV technique are 
not studied for oscillations intensively enough to deliver masses. This situation 
will hopefully improve with Transiting Exoplanet Survey Satellite (TESS; \citealt{Ricker2014}) 
or PLATO~2.0 \citep{Rauer2014}, especially with the new, more precise parallaxes 
from GAIA \citep{Gilmore1998, Perryman2001}. 

It is not surprising then, 
that masses of evolved stars studied in planet searches are very uncertain. 
If  systematic effects are present in addition to large 
uncertainties in stellar mass estimates, the 
problem may be even more severe as masses for some types of stars may be 
systematically incorrect. In this context it is important to note that one of 
the most striking features of known planetary systems around evolved stars is a 
significant stellar and companion mass increase for more evolved stars 
\citep{Niedzielski2015b}.
An average dwarf\footnote{After \cite{Niedzielski2015b} in the following  we will, for simplicity,  call 
dwarfs stars with $\log g =4.5\pm0.5$, 
subgiants:  $\log g=3.5\pm0.5$, 
giants:  $\log g=2.5\pm0.5$ 
and bright giants: $\log g=1.5\pm0.5$} 
with a planetary system  is usually a solar-mass, F or G spectral 
type star. An average subgiant known to poses a planetary system is already 
a star with a mass of $\approx1.5\Msun$  (MS spectral type A-F),
and an average giant with planets (MS A-type star) is almost twice as massive as a 
dwarf  ($5-10\sigma$ difference). 

On the other hand among the known stars with planets an average giant hosts a 
companion about twice as massive as a dwarf, and a bright 
giant's companion is 3 times more massive, on average. 
That puts  companions  to bright giants, on average, at the brown dwarf -- 
planet borderline (see also \citealt{2013A&A...555A..87M}).
It is rather obvious that such an increase a giant planet mass cannot be explained 
by accretion during its host's red giant branch evolution   \citep{1998Icar..134..303D}. 

Stellar mass is not expected to increase during MS and the subgiant branch evolution. 
Therefore stellar masses of evolved stars in planet searches are sometimes considered 
overestimated \citep{Lloyd2011, Lloyd2013, SchlaufmanWinn2013}.
 This is a very important issue as uncertainties in stellar masses immediately lead 
 to uncertainties in planetary masses and in a consequence make statistical 
 considerations of exoplanets more difficult. \cite{Sousa2015} already considered 
 in more detail masses of hosts of known planetary systems and found some of them 
 overestimated. Here we will present a more general approach to this problem.

With this paper we continue a series dedicated to detailed description of the complete 
sample of $\sim1000$ stars studies within PTPS. This project is performed with 
the Hobby-Eberly Telescope (HET, \citealt{Ramsey1998}) and devoted to planets in evolved planetary
 systems. So far 20 planets in 17 systems have been found 
 \citep{Nie2007,Nie2009a,Nie2009b,Gettel2012a, Gettel2012b,Adamow2012,Nowak2013,Niedzielski2015a,Niedzielski2015b}. 

 In the first paper of this series the red giant clump (RGC) sample was presented (\citealt{ZIELinski} - hereafter Paper~I).
 In the second one we presented Li abundances in that sample (\citealt{Adamow2014} - Paper~II).  
The purpose of the present paper is to deliver physical parameters, such 
as effective temperatures ($\Teff$), stellar gravitational accelerations
($\log g$), microturbulence velocities ($\vt$) and metallicities  ([Fe/H])  for 
455~GK-type stars, presumably subgiants and red giants observed within PTPS survey. The 
atmospheric parameters, together with existing photometric data and parallaxes 
(when available) will allow us to estimate stellar masses  ($M/\Msun$), radii 
($R/\Rsun$) and ages. 
We will also discuss the complete PTPS evolved stars sample 
and stellar masses in our and several other planet searches.

The scope of the paper is the following: in  Section \ref{targets}, we describe the sample and the observational 
material to be used in the analysis. 
The spectroscopic analysis of collected data is described in more detail  in Section \ref{s_analysis}. 
In Section \ref{LMAR}  we present stellar integrated parameters: masses, luminosities, ages and radii.
Section \ref{PTPS_sample} contains a short description of the complete PTPS evolved stars sample while in 
Sections \ref{Sec_others} we compare the sample to other samples of evolved stars searched for planets, 
for which basic data are available in the literature.
In Section  \ref{discussion} a discussion of results is presented together with a more detailed analysis 
of the origin of planetary masses increase apparent in evolved planetary systems.
Section \ref{conclusions} contains short conclusions of the paper.

\section{Targets selection and observations}\label{targets}

Spectroscopic observations presented here were made  with the HET  and its High Resolution Spectrograph (HRS, 
\citealt{Tull1998}) in the queue scheduled mode \citep{Shetrone2007}. 
The spectrograph was used in the R=60~000 resolution  mode
and it was fed with a 2~arcsec fiber. 
The configuration and observing procedure were identical to those described 
in Paper~I. 

Collected spectra consist of 46 "blue" echelle orders (407 - 592~nm) and 24 
"red" orders (602 - 784~nm). Data reduction was done with a pipeline based on
IRAF\footnote{IRAF is distributed by the National Optical Astronomy 
Observatories, which are operated by the Association of Universities for 
Research in Astronomy, Inc., under cooperative agreement with the National 
Science Foundation.} tasks (flat fielding, wavelength calibration and normalization to 
continuum). The signal to noise was typically better than 200 per resolution 
element. For every star at least one so-called GC0 spectrum is available,
 which is a spectrum obtained without a I$_2$ gas cell 
inserted into optical path, and a series of GC1 spectra, obtained with the gas cell inserted.
The observational material used in this paper are the best quality GC0 spectra and all available GC1 spectra for
 a sample of 455 subgiant and giant stars.
The sample includes 11 SB1 systems. 

\begin{figure}
   \centering
   \includegraphics{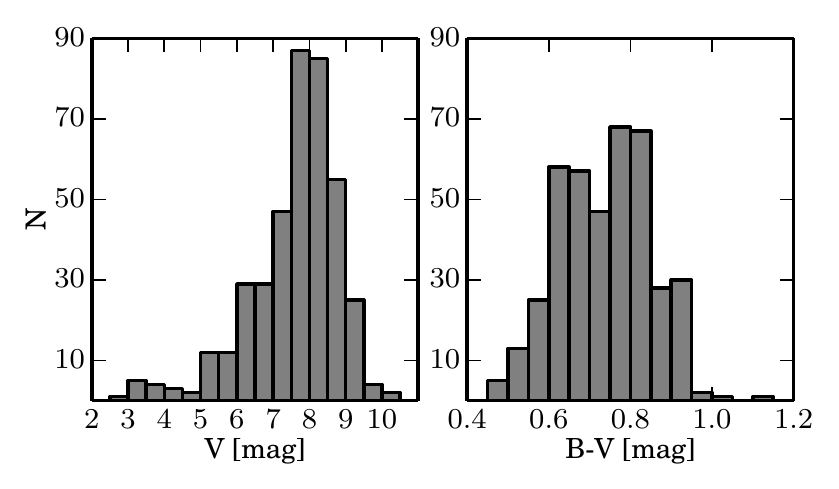}
      \caption{Histograms of the apparent magnitudes in the V band (left panel) 
                   and $(B-V)$ (right panel) for 402 stars observed within PTPS. }
   \label{fig-obs}
\end{figure}

The sample was designed as a {\it blind} extension of the RGC 
 sample that had been observed with HET (see Paper~I)
and consists of a set of  field
stars that meet several requirements.
One of them is location on Herztsprung-Russell Diagram (HRD)
that corresponds to subgiants and giants. To identify those
objects we used photometric data and parallaxes from Hipparcos and Tycho catalogs,
choosing stars with B-V=0.55--0.95, located in expected $M_V$ range. 
For efficient use of HET/HRS observing time, selected stars are
 randomly distributed over the HET field of view ($\delta=-10^{\circ} 20'$ to $+71^{\circ} 40'$).
 Adopted observing strategy  puts a limit for stellar brightness of 
 observed stars. In case of PTPS, observed stars should be brighter than $10.5^{mag}$
 and this threshold was also applied during the selection. 
The complete list of selected targets is presented in Table \ref{tab-res} (see also Fig. \ref{fig-obs}).

\addtocounter{table}{1}

\section{Spectroscopic analysis}\label{s_analysis}

The spectroscopic analysis included a check for presence of stellar companions with a 
cross-correlation technique, absolute radial RV measurements,  
atmospheric parameters determinations with an LTE analysis of Fe~I and II 
lines, and rotation velocities estimates. 

This  approach was proved to be robust.  For three objects from Paper~I, HD 102272, BD+20 2457 and BD+48 738,
practically identical (within $1\sigma$) atmospheric parameters  were obtained  by \cite{Mortier2013} 
(except [Fe/H] for BD+20 2457 that was found $\sim 3\sigma$ lower).  Agreement within $1\sigma$ was also obtained for parameters of
 another five stars analyzed by \cite{Sousa2015}:  HD~17092, HD~240210,  HD~240237, HD~96127 and HD~219415 .

\subsection {CCF analysis \label{CCF}}

To construct the cross-correlation functions (CCFs) we correlated 
all available GC1 stellar spectra for every stars with a numerical mask consisting of 1 and 0 
value points, after cleaning the spectra from the I2 lines, using the ALICE code \citep{Nowak2012phd, Nowak2013}. 
The non-zero points correspond to the positions of 
300 non-blended, isolated stellar absorption lines at zero velocity, present in 
a synthetic ATLAS9 \citep{Kur1993a} spectrum of a K2 star.  We used only the first 17 orders 
of the ''blue'' spectra. They  are free from telluric lines and corrected in the 
wavelength scale using the I$_2$ lines. The CCF was computed step by step for each 
velocity point in a single order. For every order the algorithm 
selected from the mask only these lines that are suitable for a given wavelength 
range. CCFs from all orders were finally added to get the final CCF for the whole 
spectrum.  

\begin{figure}
   \centering
   \includegraphics{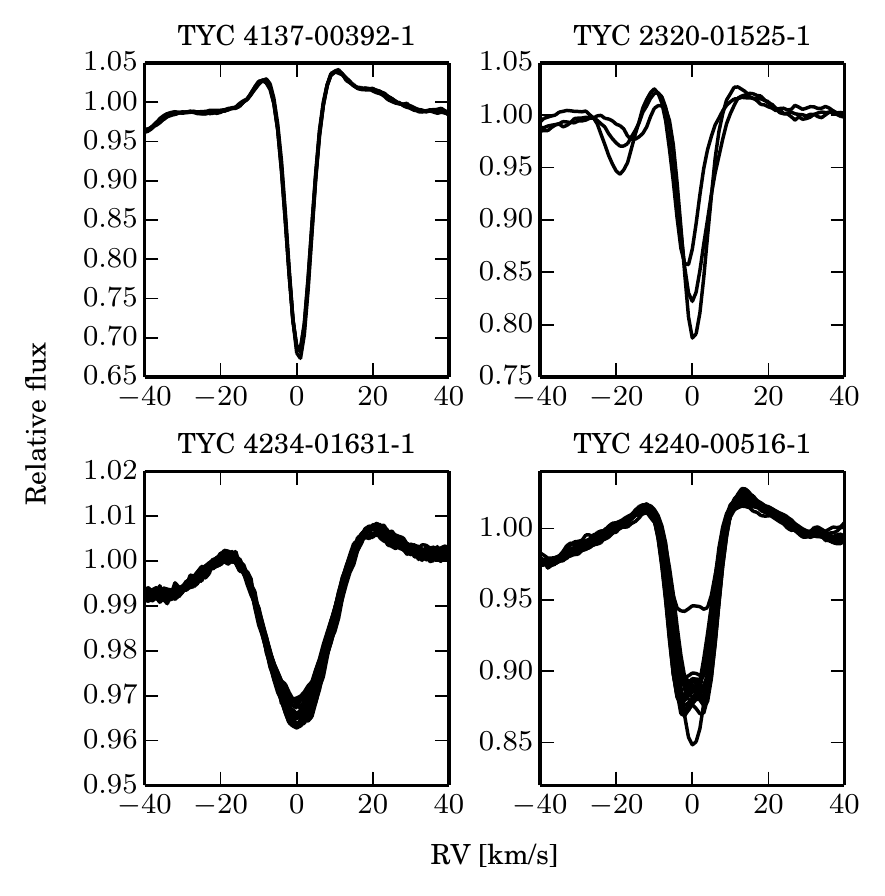}
   \caption{Sample of CCF for four objects: top left - single star or SB1; top right - SB2; 
                 bottom left - week line (low metallicity or fast rotator); both right: variable CCF/SB2.}
    \label{fig-ccf}
\end{figure}

The shape of the CCF and its variation in the series of available spectra were used to identify spectroscopic 
binaries with resolved spectral line systems (SB2), objects with variable CCF  and 
stars  with  flat CCF (fast rotators or low metallicity stars). Typical CCFs for various cases are presented
in Fig.~\ref{fig-ccf}.
A summary of results of CCF analysis  is presented in Table 
\ref{table-rejected}. 

Due to the nature of star selection for the presented sample, 
after the preliminary analysis some objects 
appeared  unsuitable for a planet search and were rejected.
Altogether 53 stars were rejected based on various conditions.
Our CCF analysis revealed a group of 25 stars with multiple CCF and 22 with unresolved 
and variable depth of CCF. A ``Spectroscopic binaries - SB2'' and ``Variable CCF/SB2 ''
label was assigned to those stars, respectively. For another group of 5 stars
 the absorption line system was too weak to
measure EW realistically. To those stars we assigned ``weak CCF'' label,
and they may be fast rotators or very low metallicity stars also unsuitable for a planet search. 
We excluded  also from further analysis one object with probably very low effective temperature.

 For those stars a detailed spectroscopic analysis was not possible as their equivalent 
widths are misleading or accompanied with huge uncertainties.
To roughly estimate atmospheric parameters for  these stars in our analysis we adopted for all of 
them $\Teff$, $\log g$ as well as initial luminosity estimates from 
\citet{Adamow2010}, i.e. values of $\Teff$ obtained from empirical calibration of 
\citet{RM2005} based on Tycho and 2MASS photometry as well as $\log g$ roughly 
estimated using the method of \citet{Bil2006} and \citet{Gel2005}. 
The results of this simplified analysis are presented in Table 
\ref{table-rejected} for completeness but they will be ignored in further analysis.

\subsection{Absolute radial velocities}

The absolute radial velocity  was measured by fitting  a Gaussian function to 
the co-added CCFs for the 17 orders of the blue spectrum of the best available 
GC0 spectrum, obtained with the ALICE code \citep{Nowak2012phd, Nowak2013}. 

\begin{figure}
   \centering
   \includegraphics{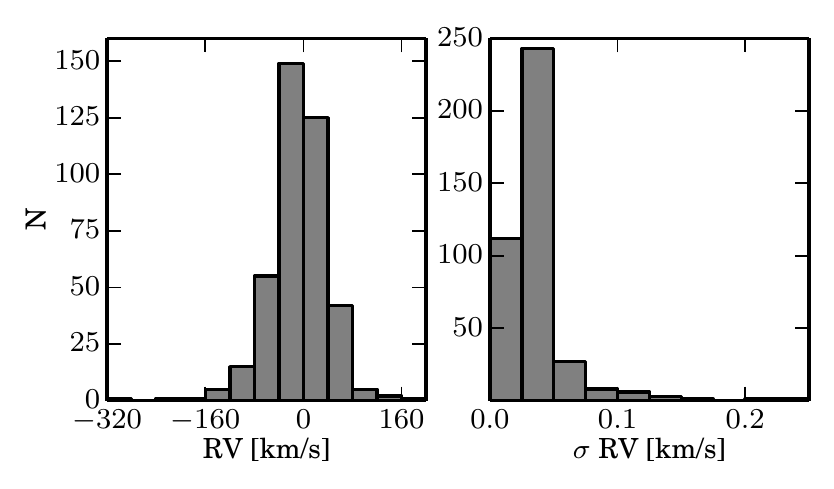}
      \caption{Histograms of the RVs obtained from the cross-correlation 
                    function (left panel) and their uncertainties (right panel) for studied 
                    PTPS stars.}
   \label{fig-rvcomp}
\end{figure}

The uncertainty  of 
the resulting RV was computed as $RMS/\sqrt{17}$ of the 17 RVs obtained  for 
each order separately. The mean standard uncertainty obtained in this way is 
$\sigma \rm{RV}_{CCF}=0.039 \kms$. However, as 
HET/HRS is neither thermally nor pressure stabilized, the RVs
are subject to seasonal variations and the actual precision is in the $2-3 \kms$ range. The 
distribution of  RVs and their uncertainties is presented in Fig.~\ref{fig-rvcomp}.

 RVs were transformed to the barycenter of the 
Solar System with the algorithm of \citet{Stum1980} for all stars in our sample 
They are presented in Table~\ref{tab-res} (column 11), together 
with the epochs of observation as modified julian date (MJD, column 12).

\subsection{Rotational velocities}

\begin{figure}
   \centering
   \includegraphics{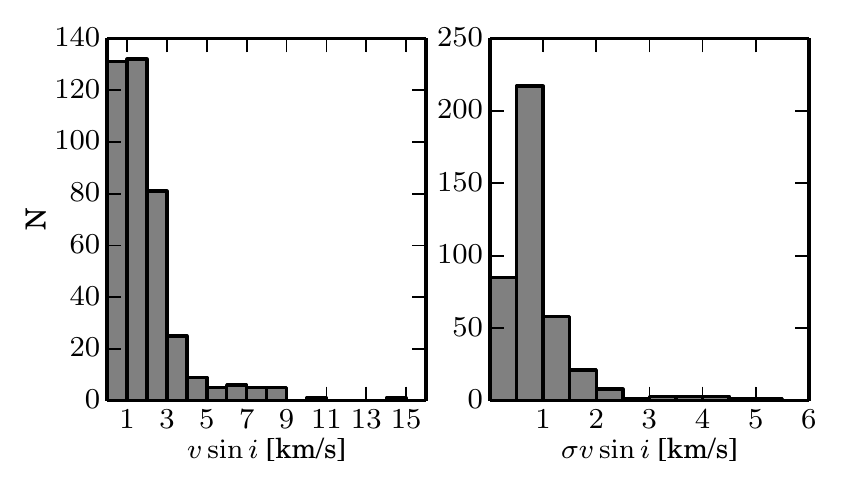}
      \caption{Histograms of the rotational velocities and its uncertainties 
                    obtained from SME  \citep{ValPisk1996} synthetic spectrum fitting.}
   \label{fig-vrot}
\end{figure}

We obtained projected rotational velocities, $v \sin i_{\star}$, 
through modeling of GC0 and ,,red''
stellar spectra with the Spectroscopy Made Easy tool (SME,  \citealt{ValPisk1996}). 
A more detailed description of this method is presented in Paper~II.
Estimated rotational velocities are presented in Table ~\ref{tab-res} (column 10).
Majority of our objects are slow rotators with $v \sin i_{\star}$ of $1 -3 \kms$. 
Only two objects show $v \sin i_{\star}$ larger than $10 \kms$: TYC~1427-00095-1 
and TYC~1515-00866-1.
Histograms of our estimates of $v \sin i_{\star}$ and its uncertainties are 
shown in Fig.~\ref{fig-vrot}. 

\subsection{Equivalent widths \label{Sect_EW}}

We used Automatic Routine for line Equivalent widths in stellar Spectra (ARES, 
\citealt{Sou2007}) to measure the  equivalent widths (EWs) of spectral lines.  
The code allowed us to automatically measure EWs for a set of 
neutral and ionized iron absorption lines and we found it more 
flexible than DAOSPEC \citep{SP2008} used in Paper~I. 
To test the consistency between ARES  and  DAOSPEC measurements, we
compared  EW's measurements of Fe lines  (with $\rm{EW}<200\rm{m}\AA$)
 for 5 stars (HD 3933, 102842, 187094, 215443, 9416) 
obtained with both tools.
We found a very good agreement between the two sets 
of measurements (the Pearson correlation coefficient r=0.943),
$\rm{EW}_{DAOSPEC} = 0.911\pm 0.017 \times \rm{EW}_{ARES} + 5.049 \pm0.852$,
with an average difference of 
$ 6.50 \pm 10.22 \rm{m}\AA$
and average ratio 
of $\rm{EW}_{ARES}/ \rm{EW}_{DAOSPEC}=  1.06 \pm 0.41$ 
We found it justified to continue with  ARES. 

The selection of spectra lines used in spectroscopic analysis may lead 
to slight variations in results \citep{Tsantaki2013,Alv2015}.
In this paper, for consistency with Paper~I, and to exploit the available spectra 
we chose the line list by \cite{Tak2005a} as the most adequate for our 
HET/HRS spectra following the results of analysis of  \cite{Adamow2015}. 
We  removed from it all lines in  regions of strong telluric lines occurrence. 
Finally we included 220 lines (200 Fe~I and 20 Fe~II) in the $4813-7855 \AA$ range
in our  spectroscopic analysis.

\subsection{Atmospheric parameters}

Having EW measurements 
we applied them as 
input data to determine the LTE atmospheric parameters of our sample stars.  
The $\Teff$, $\log g$, $\vt$ and [Fe/H]  (defined in standard manner as [Fe/H]$=\log 
(N_{\rm{Fe}}/N_{\rm{H}})-\log (N_{\rm{Fe}}/N_{\rm{H}})_{\odot}$), 
were obtained with TGVIT code  developed by \citet{Tak2002a, Tak2005a}.  
See Paper~I for more detailed description of our 
implementation of the code. 
For [Fe/H] calculations, we adopted the $A(\rm{Fe})$ solar value 
of 7.50~dex \citep{Kur1993a,Hol1991}.
For 402 objects we obtained converged solutions for each of stellar parameters 
typically in 10 or less iterations.

To test presented methodology of  atmospheric parameters determination, we 
applied it to Arcturus. We acquired several spectra for this object within PTPS 
and we used the one of the best quality (S/N=420). 
The atmospheric parameters we obtained: $\Teff = 4254\pm 20 \rm{K}$, $\log g = 
1.61 \pm 0.08$, $\vt = 1.54 \pm 0.07$ and [Fe/H]$ = -0.61 \pm 0.08$. 
are in very good agreement with those from \citet{Ramarc2011}: $\Teff = 4286 \pm 
30 \rm{K}$, $\log g = 1.66 \pm  0.5$, 
and [Fe/H]$= -0.52 \pm 0.04$, and from \citet{Mararc1977}: $\Teff = 4300 \pm 90 
\rm{K}$, $\log g = 1.74 \pm 0.2$, $\vt = 1.70$ and [Fe/H]$= -0.51\pm 0.08$ (see 
Sect \ref{errors} for a comment on atmospheric parameters uncertainty).
We are therefore confident that the applied methodology is correct.

\begin{figure}
   \centering
   \includegraphics{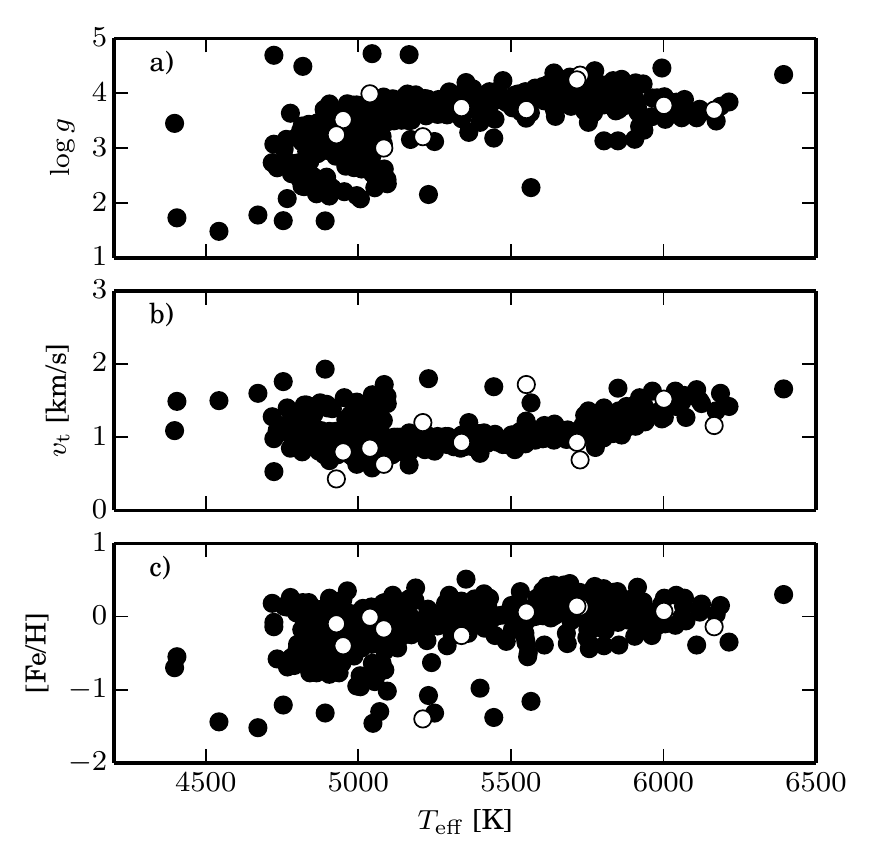}
   \caption{Relations between $\Teff$, $\log g$, $\vt$ and [Fe/H] for 402 stars with 
                 complete spectroscopic analysis. Results for 11 SB1 are  marked as open circles.}
         \label{fig-tgmv}
\end{figure}

\begin{figure*}
   \centering
   \includegraphics{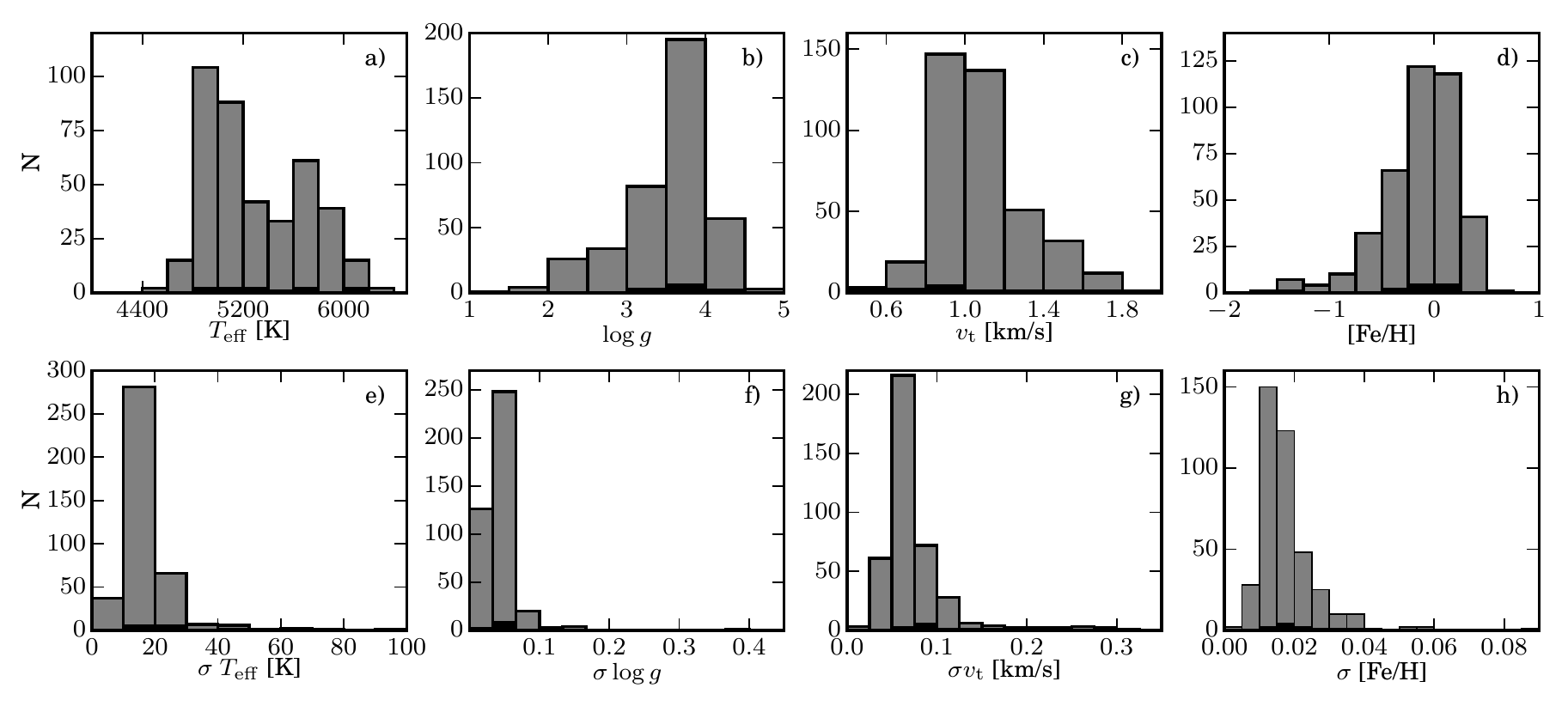}
   \caption{Histograms of   of $\Teff$, $\log g$, $\vt$ and [Fe/H] obtained
	        for 402 stars with complete spectroscopic analysis. Results for 11 SB1 are
		marked in black.}
   \label{fig-histsigpar}
 \end{figure*}

 \begin{figure}
   \centering
   \includegraphics{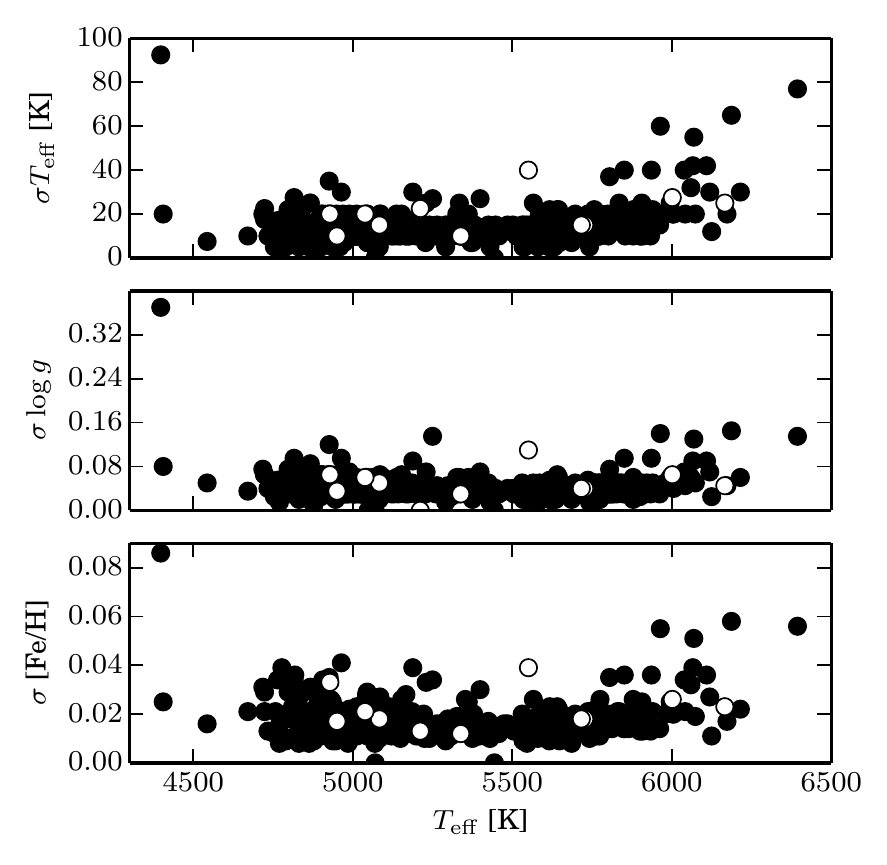}
      \caption{Relations between intrinsic uncertainties  of $\sigma\Teff$ (top panel), 
                    $\sigma\log g$ (middle panel) and $\sigma$[Fe/H] (bottom panel) 
                   vs. $\Teff$  for 402 stars. Results for 11 SB1 are marked as open circles.}
    \label{fig-sigpar}
\end{figure}

 The results of our determinations of atmospheric parameters for 402~GK stars are
 presented in Table ~\ref{tab-res} (columns 6-9), where the numerical values of
 $\Teff$, $\log g$, $\vt$ and [Fe/H] are presented together with their {
 intrinsic uncertainties}.

The values of resulting atmospheric parameters 
were found to stay generally within the range of the TGVIT model grids.

 Fig.~\ref{fig-tgmv} presents relations between atmospheric parameters 
presented in Table ~\ref{tab-res}. We note that results obtained  
for single line spectroscopic binaries (SB1) agree very well with those for apparently 
single stars. 

The effective temperature  $\Teff$ is in range between 4398~K and 6394~K with a 
median value at 5167~K. The distribution of $\Teff$, presented in 
Fig.~\ref{fig-histsigpar}a, shows two maxima, one for giants (cooler) and one for 
subgiants (warmer), which are the two most numerous groups of stars in the presented  
sample. The intrinsic uncertainty distribution in $\Teff$ is presented in 
Fig.~\ref{fig-histsigpar}e.

The gravitational accelerations $\log g$ for the presented sample  range between
1.49 and 4.72 with the median of 3.65. The distribution of $\log g$ is presented
in Fig.~\ref{fig-histsigpar}b. We found that 5 stars have $\log g<2.0$, 60 have $3.0>\log g\geq 2.0$, 
 277 have $4.0>\log g\geq 3.0$ and  60 stars have $\log g \geq 4.0$.
We can see that majority of our stars (194)  have $\log g$ of $3.5-4.0$ making
them generally subgiants. The uncertainty distribution in $\log g$ is
presented in Fig.~\ref{fig-histsigpar}f.

The microturbulence velocity, $\vt$, reaches values from $0.43 \kms$ to $1.93 \kms$ 
and has the median at $1.03 \kms$. Most of our stars have $\vt$
between $0.8 \kms$ and $1.2 \kms$ (Fig.~\ref{fig-histsigpar}c).
The uncertainty distribution in $\vt$ is presented in Fig.~\ref{fig-histsigpar}g.

The metallicity of stars in the presented  sample, [Fe/H], stays within $-1.52$ 
to $+0.51$ limit with the median value at $-0.08$. Most of our objects have the 
[Fe/H] in the range of $-0.25$ to $0.25$. Fig.~\ref{fig-histsigpar}d presents the 
distribution and we can see that our stars are mostly less metal abundant than 
the Sun. The uncertainty distribution in [Fe/H] is presented in 
Fig.~\ref{fig-histsigpar}h.

\subsection{Atmospheric parameters uncertainty estimates \label{errors}}

The mean intrinsic, i.e. delivered by TGVIT,  uncertainties of our 
determinations are: $\sigma\Teff=15$~K, $\sigma \log g=0.04$, 
$\sigma\vt=0.07\kms$ and $\sigma$[Fe/H]~$=0.02$.  
Fig.~\ref{fig-sigpar} shows the intrinsic uncertainties of these three 
parameters as a function of $\Teff$. No correlation exists 
between uncertainties and the obtained parameter values for any of the 
atmospheric parameters in the wide range of  $\Teff$  between 4500~K  -- 6000~K.
The scatter of the uncertainties is uniformly 
distributed over the whole range of resulting parameters.
We note, however,  that for stars with $\Teff$ below 4500~K  or above 6000~K the 
uncertainties are slightly higher, especially in $\Teff$ and [Fe/H].  

To test the impact of EWs shift between DAOSPEC and ARES measurements 
presented in Section \ref{Sect_EW} we calculated stellar atmospheric parameters 
for the same 5 five stars using DAOSPEC and ARES EWs separately. We obtained 
agreement within 1$\sigma$ intrinsic in all parameters except microturbulence velocity, 
in which the results differ by 2-3 intrinsic $\sigma$. We are allowed to assume, therefore, 
that our results presented here are consistent with those of Paper~I.

All these intrinsic uncertainties, except [Fe/H], are numerical uncertainties 
resulting from the iterative procedures of TGVIT, representing actually a 
goodness of fit only.
The intrinsic uncertainties in metallicities are estimated in TGVIT from the 
actual Fe abundance distribution as the standard deviation of the mean.
A detailed  comparison with \citet{Sou1998} and \citet{But2006} presented in 
Paper~I suggests that our intrinsic uncertainties in $\Teff$, $\log g$ and $\vt$  are underestimated by a factor of 2-3. 
In Table \ref{tab-res} we present, however, the intrinsic uncertainties from 
TGVIT allowing for future more detailed uncertainty analysis.

Both atmospheric parameters and their uncertainties for SB1 stars in the presented sample
fit well the general trends for single stars (Fig. \ref{fig-tgmv}, \ref{fig-sigpar}) 
and therefore in the following we will address single and SB1 stars together.

\begin{figure}
   \centering
   \includegraphics{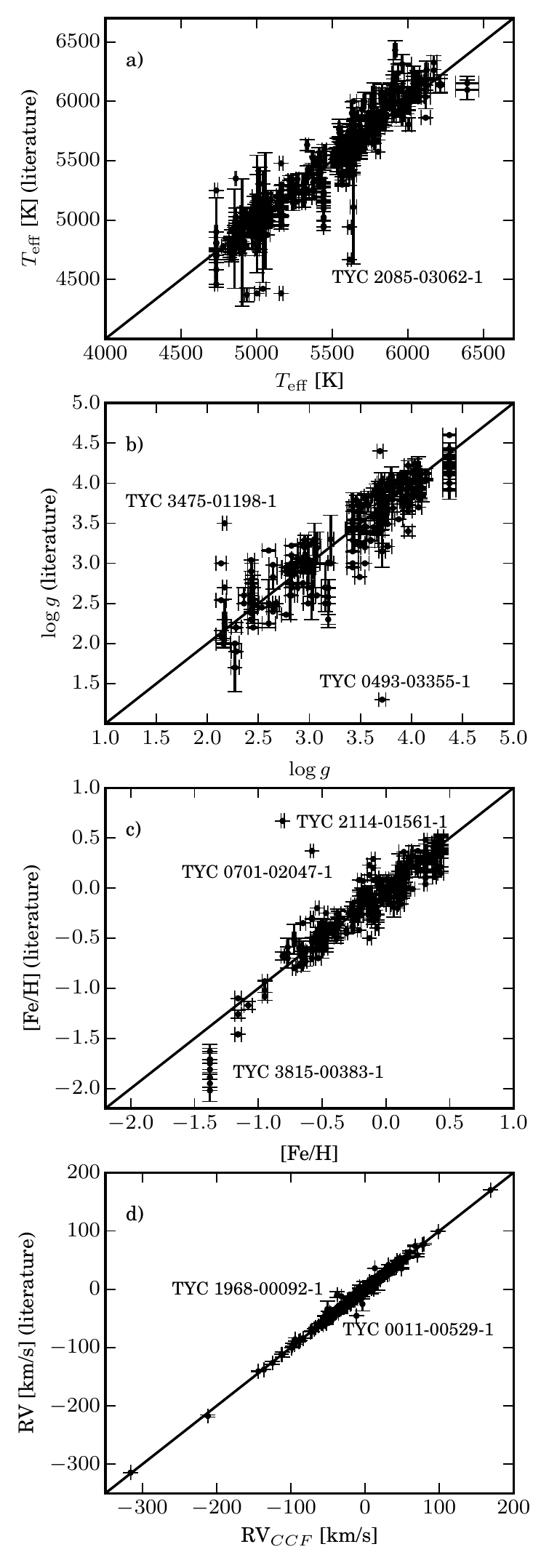}\vspace{-0cm}
      \caption{Comparisons between our $\Teff$ ,$\log g$, [Fe/H] and literature 
      values from PASTEL catalogue 
\citep{PASTcal}. Radial velocities derived from our cross-correlation analysis are
      compared with measurements available in SIMBAD. The RVs uncertainties are 
      presented if available. The solid lines present the one to one relations.}
         \label{fig-pastel}
\end{figure}

\subsection{Comparison with literature }
Atmospheric parameters for 272 stars discussed here were determined for the first time. 

In our attempt to compare  our results with those available in PASTEL catalogue 
\citep{PASTcal} we  found 
130 stars in common, many with multiple records in PASTEL resulting sometimes in large scatter. 
After removing from the comparison the objects with most discrepant values (labeled in Fig. \ref{fig-pastel}) 
good  agreement was found in the case of all parameters.

For $\Teff$ (444 measurements in Pastel)- the Pearson correlation 
coefficient is r=0.96 (Fig.~\ref{fig-pastel}a), mean difference between measurements 
$\delta=87\pm86$~K and average our to their result ratio is $\rho=1.00\pm0.02$.

For $\log g$ (244 records in Pastel) we also found very good agreement with r=0.92, 
$\delta=0.19\pm0.15$ and $\rho=1.02\pm0.08$ (Fig.~\ref{fig-pastel}b).

In the case of   [Fe/H] (248 records in Pastel),  we found r=0.95, 
$\delta=0.09\pm0.08$ and $\rho=0.75\pm2.47$ (Fig.~\ref{fig-pastel}c). 
Technically this result suggests that our [Fe/H] determinations are systematically
lower than those present in Pastel. That result is, however, misleading 
and caused by a scatter in results for $|$[Fe/H]$|<$ 0.1, usually actually
 consistent within uncertainties. For 181 records with $|$[Fe/H]$|>$ 0.1 we found $\rho=0.93\pm0.5$.

In Fig.~\ref{fig-pastel}d  the RVs obtained here with the CCF technique are 
compared with literature data for 279 objects  included 
in SIMBAD\footnote{The SIMBAD astronomical database is operated at CDS, 
Strasbourg, France.} database. We found r=0.97, $\delta=1.8\pm3.2 \kms$, and $\rho=0.98\pm0.92$).

No systematic effects were found. Our results agree  with those obtained by other authors 
within estimated uncertainties.

\section{Luminosities, masses, ages and radii}\label{LMAR}

We  estimated intrinsic color index $(B-V)_{0}$ and bolometric corrections 
$BC_{\rm{V}}$ for our stars using empirical calibrations by \citet{Alo1999}. 
For 374 stars with available Hipparcos parallaxes \citep{vanLeeuwen2007}  assuming the 
standard interstellar reddening with 
total to selective extincion  ratio
$R_{V}=3.1$ \citep{Rie1985}  
luminosities were directly calculated. These luminosities were used in further analysis.

\begin{figure*}
   \centering
   \includegraphics{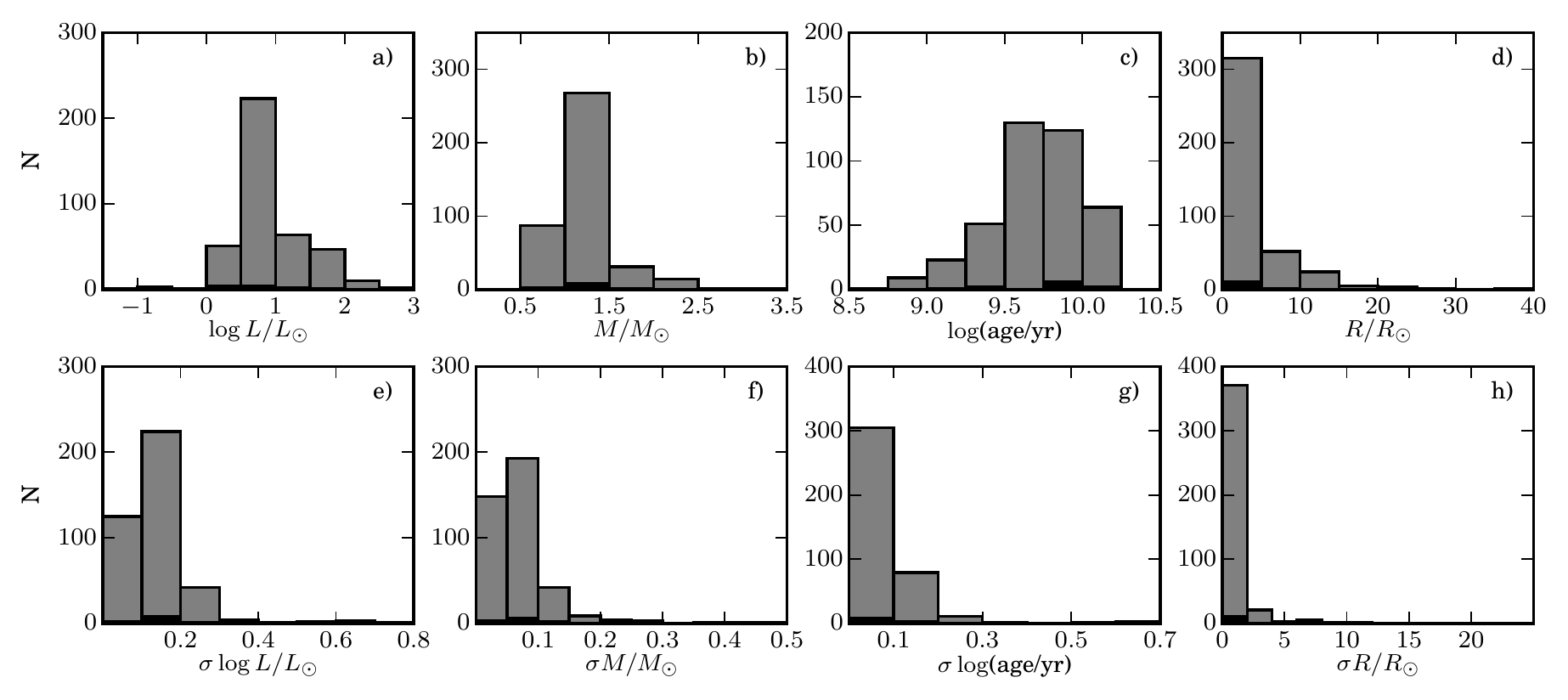}
      \caption{Distributions of $\log L/L_{\odot}$, $M/\Msun$, log(age/yr) and $R/\Rsun$ for 402 stars 
      with complete spectroscopic analysis (panels a-d). The histograms for 
      uncertainty estimates are 
      presented in panels e-h. Results for 11 SB1 are marked in black.}
         \label{fig-abs}
\end{figure*}

\subsection{Stellar masses and ages}

Stellar masses $M/\Msun$ and  estimates of ages were obtained with the  
\cite{Adamczyk2015} modification of the Bayesian method based on 
\cite{JorgensenLindegren2005} formalism, modified by \cite{daSilva2006}  
to avoid statistical biases and to take uncertainty estimates of  observed quantities 
into consideration. We adopted theoretical stellar models  from \cite{Bressan2012} 
and used isochrones with  Z= 0.0001, 0.0004, 0.0008, 0.001, 0.002, 0.004, 
0.006, 0.008, 0.01, 0.0152, 0.02, 0.025, 0.03, 0.04, 0.05, 0.06 and 0.008 
interval in $\log\mathrm{(age/yr)}$. The adopted solar distribution of heavy elements 
corresponds to  Sun's metallicity $Z \simeq 0.0152$ \citep{Caffau2011}. The 
helium abundance for given metallicity was obtained from relation 
$Y=0.2485+1.78Z$ \cite{Bressan2012}.  
For majority of our targets with Hipparcos parallaxes  available we used 
$\Teff$,  $\log g$ , [Fe/H]  and luminosity as input parameters for the Bayesian analysis.
In case of stars with no reliable parallax we were able to obtain  from the Bayesian analysis 
also an estimates of luminosity.

A detailed description of that approach is presented in 
\cite{Adamczyk2015}.

The resulting masses range from $0.52 \Msun$ to $3.21\Msun$ as the histogram 
of $M/\Msun$ for all our stars presents (Fig.~\ref{fig-abs}b). We can see that 
majority of stars have masses up to $1.19\Msun$ (median). However, there is a significant 
number (17 or $\sim 4.2\%$) of stars which fall in the intermediate-mass range 
$2\Msun \le M \le 3.21\Msun$. Table~\ref{tab-res} (column 17) presents the final 
adopted masses.

The uncertainties of stellar masses obtained by comparison with theoretical 
isochrones depend on the accuracy of determination of star's location 
in the [$\log L/L_{\odot}$, $\log g$, $\log\Teff$] space. 

The main source of uncertainty in mass is the parallax, or luminosity, and its
uncertainties. In the case of stars, for which the parallax  was more precise 
the simplified treatment of metallicity may introduce additional 
uncertainty in mass through the choice of the metallicity model and then 
isochrone.  
We found that for the  whole sample the mean 
uncertainty for stellar mass is $\sigma M/\Msun=0.07$. However, uncertainties 
may become much larger 
for very confusing isochrones at 
lower temperatures ($\Teff \le 4500$~K). We also note that 
for stars from the RGC, which may undergo stellar mass-loss 
the masses obtained this way are  the upper limits as in \cite{Bressan2012} 
the stellar mass lost due to mass loss is added while constructing 
the isochrones and the MS masses are given. 

Estimated stellar ages are presented in Table~\ref{tab-res} (column 19) and in (Fig.~\ref{fig-abs}c).
We found that typical stars from our sample have $\log\mathrm{(age/yr)}$ between 9.5 and 10 and have mean 
uncertainties around 0.09.

\subsection{Stellar luminosities}

For 28 stars with no parallaxes (or $\sigma_\pi > \pi$ ) the luminosity was  
 adopted from fits to the isochrones, corresponding to determined stellar mass.
In Table~\ref{tab-res} (column 15) the final adopted luminosity is provided.  
The $\log L/L_{\odot}$ ranges from -1.25 to 2.84 with the median of 0.76 (Fig.~\ref{fig-abs}a). 

\begin{figure*}
   \centering
   \includegraphics{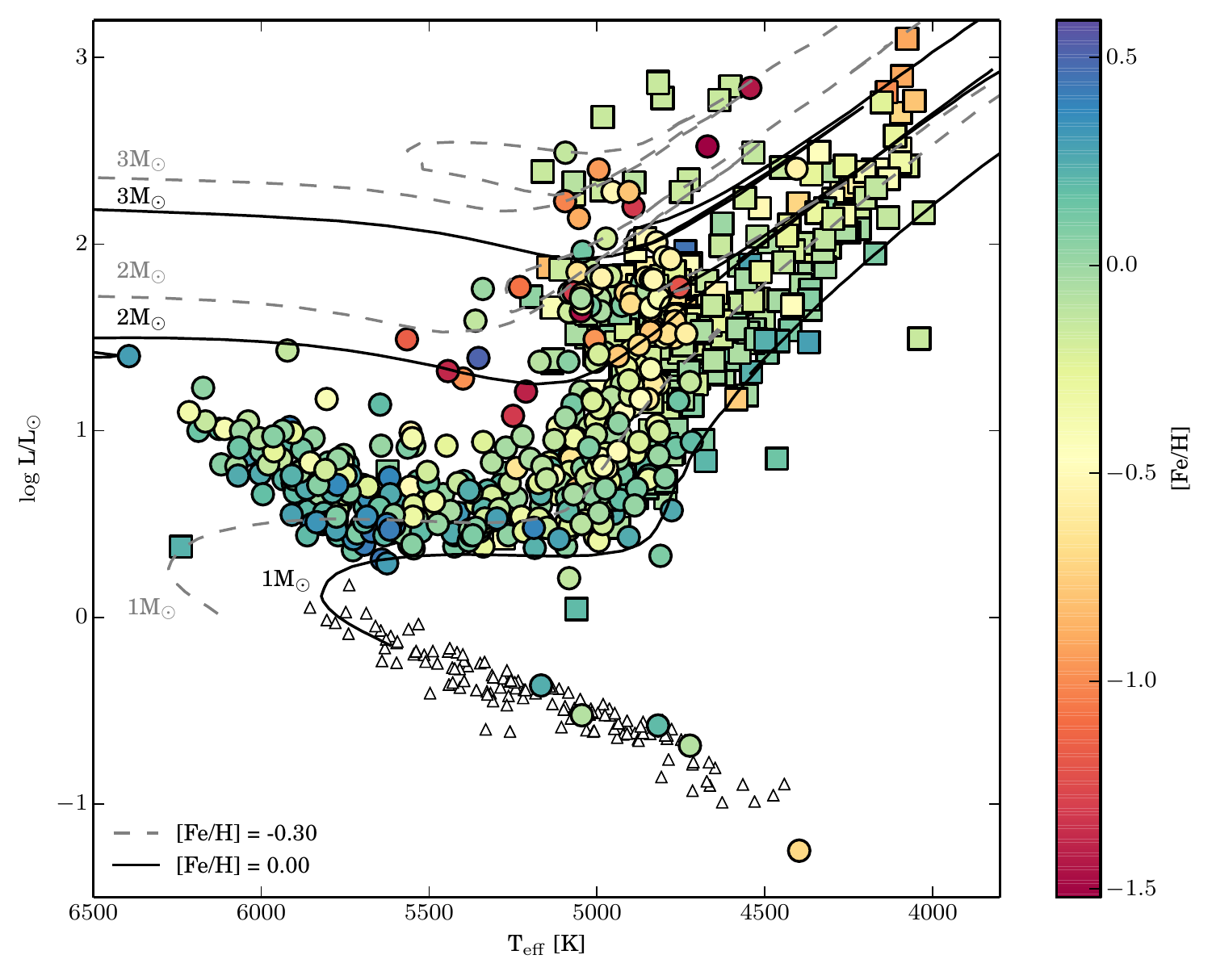}
   \caption{The H-R diagram for the complete  PTPS sample of evolved stars. 
                 The stars presented in this paper are displayed as circles, and  objects discussed in Paper~I  
                 as squares.  We also present preliminary results for dwarfs (open triangles, Deka-Szymankiewicz et al. in prep.), 
                 included also in PTPS, for completeness.  Theoretical evolutionary tracks \citep{Bertelli2008,Bertelli2009} 
                 are presented for stellar masses of $1-3\Msun$ and two metallicities: the solid 
                 and dashed lines correspond to [Fe/H] $=-0.3$ and 0.3 respectively. 
                 Stellar metallicity is color-coded.}
   \label{fig-hrd-ptps}
\end{figure*}

For stars with luminosities calculated from Hipparcos parallaxes luminosity 
uncertainties were calculated using the standard exact 
differential law for calculations based on the available $\pi$ and photometry uncertainties. 
For Bayesian estimates  they were estimated 
as dispersion of respective PDF (Fig.~\ref{fig-hrd-ptps}) for the proper stellar mass.  
The average value of luminosity uncertainty for the whole sample is $\sigma\log L/L_{\odot}=0.14$.

\begin{figure}
   \centering
   \includegraphics{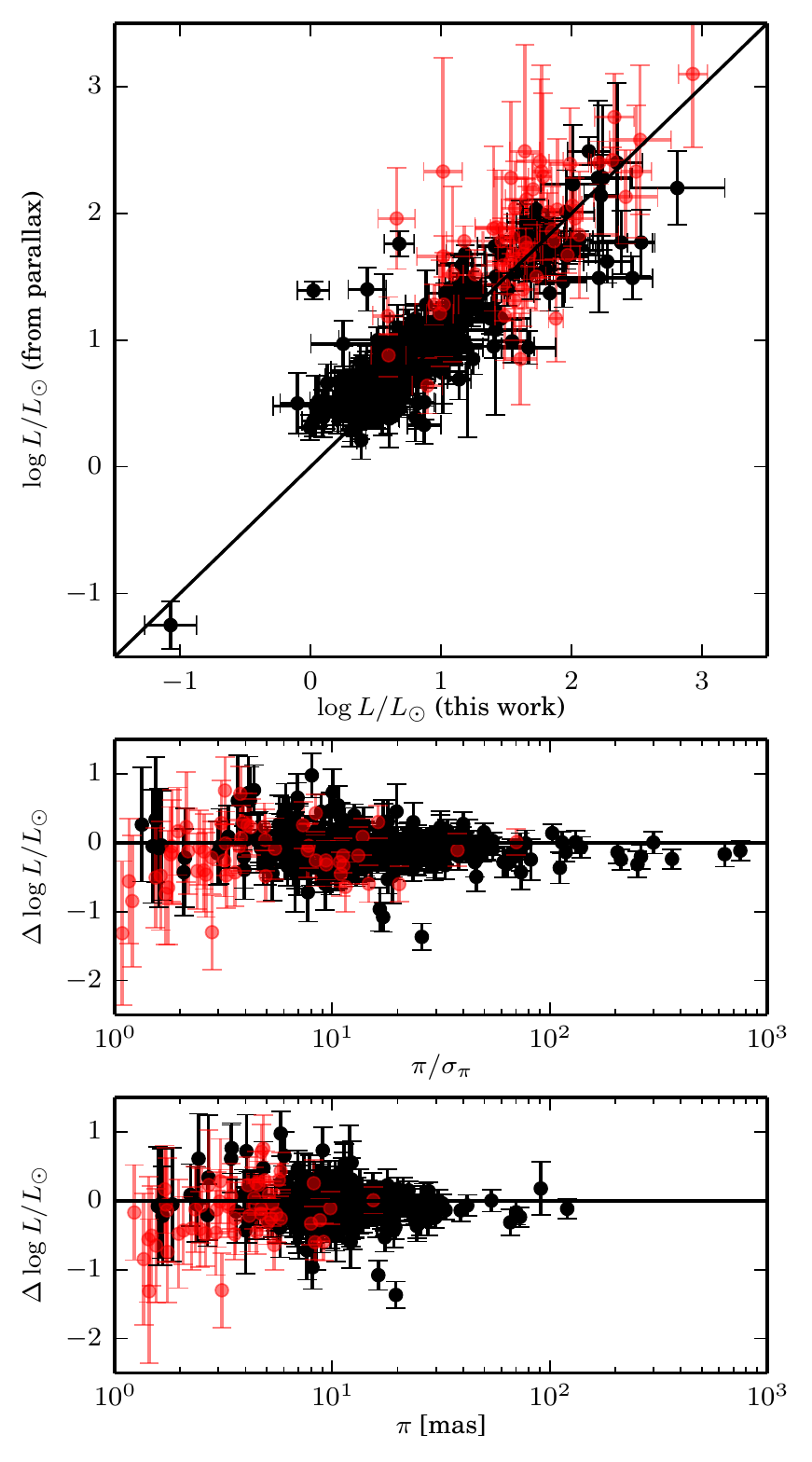}
   \caption{Comparison of stellar luminosities obtained from Hipparcos parallaxes 
                and from our Bayesian analysis. The stars presented here are represented 
                by black symbols while those of Paper I by red ones. The solid line presents the one to one relation.}
   \label{fig-LL}
\end{figure}

In Fig.~\ref{fig-LL} we present a comparison of stellar luminosities obtained 
from Hipparcos parallaxes ($\sigma_\pi < \pi$ )  and from our Bayesian analysis based on atmospheric
parameters only. A general agreement is clear although for stars with very small 
parallaxes ($\pi \lesssim10$~mas) both estimates may vary substantially. In Fig.~\ref{fig-LL} 
we also present (in red) luminosities obtained with the two methods for the Paper~I 
stars by \cite{Adamczyk2015}. That sample contains more distant stars 
for which luminosity is certainly less reliable.

\subsection{Stellar radii}

With either spectroscopic $\log g$ and adopted stellar mass or spectroscopic 
$\Teff$ and adopted luminosities, we were able to calculate  stellar radii (see 
\citealt{Adamczyk2015} for more details). Although for several stars the 
radii obtained from those two sets of parameters differ significantly, the general 
agreement between both $R/\Rsun$ estimates is good (r=0.93).
The maximum uncertainties were estimated in both approaches by application of 
the standard exact differential law. 

As it can be seen from Fig.~\ref{fig-abs}d, the radii range from $0.66\Rsun$ to 
$36.04\Rsun$. Most of our stars have radius of about $0.66-4\Rsun$. Median and 
mean  $R/\Rsun$ was found to be 2.76 and 4.24, respectively. 
On average the precision is $\sigma R/\Rsun=0.19$.
The adopted stellar radii are presented in Table~\ref{tab-res} (column18).

The ranges of adopted uncertainties of radii as well as luminosities, masses and ages 
are depicted in Fig.~\ref{fig-abs} (panels e-h). 

\section{The PTPS evolved stars sample}\label{PTPS_sample}

\begin{figure}
   \centering
   \includegraphics{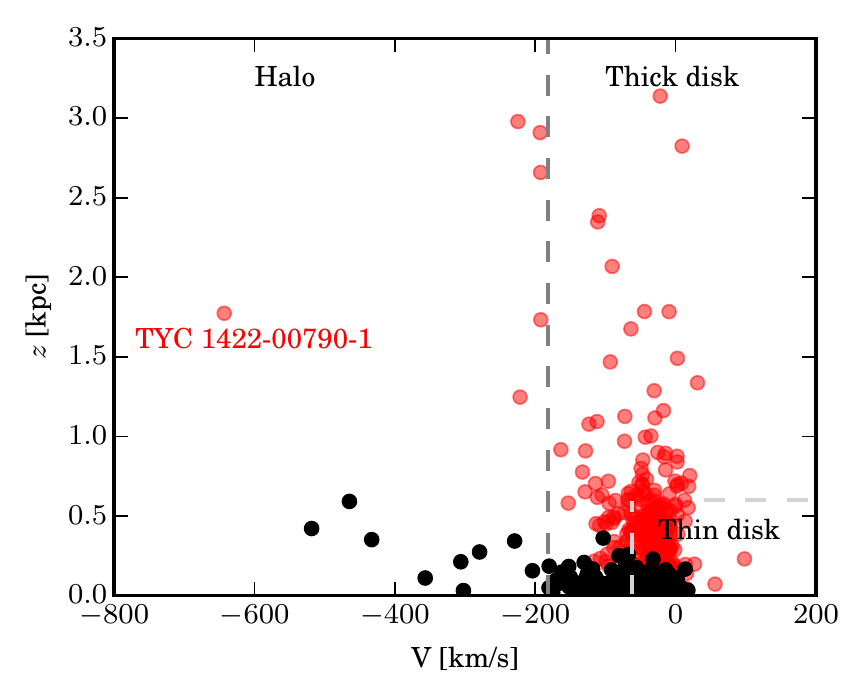}
   \caption{Galactic rotational velocities of stars with respect to the 
                Galactic Center  versus distances from the galactic plane $z$. 
                The stars presented here are represented by black symbols while those of Paper~I by red ones.}
   \label{fig-locgal}
\end{figure}

The sample of 402 stars presented in this paper, together with  342 stars from 
the  RGC sample discussed in Paper~I  with revised, integrated 
parameters presented in \cite{Adamczyk2015}, constitute the evolved stars sample of 
PTPS. For all 744 stars atmospheric parameters as well as masses, 
luminosities, ages  and radii were obtained in a uniform way. Both 
subsamples, in spite of differences resulting from their definitions  are complementary. The main 
difference between them is that 
giants and distant bright giants are mostly present among Paper~I   sample, while subgiants are
highly abundant in the sample presented here. Another difference is the amount 
of stars with Hipparcos parallaxes available. Contrary to the sample presented 
in Paper~I, for most of stars discussed in the present paper parallaxes 
are available. As a result,  luminosity, masses and ages estimates presented in this paper are presumably more 
precise (cf. \citealt{Adamczyk2015}). 
In Fig.~\ref{fig-hrd-ptps} we present the HRD for the complete evolved sample of the PTPS stars, together with a subsample of dwarfs, not discussed in this paper.

We applied \citet{JohSod1987} approach to constrain space velocity components 
(U,V,W)  for the PTPS evolved stars sample. Following \citet{Ibuk2002} criteria, we 
assigned our objects to galactic populations. 
Stars from thin disc, thick disc and halo have $V> 
-62 \kms$, $-182 \le V <  -62 \kms$ and $V<-182 \kms$, respectively. Objects 
with $V > -62\kms$   but with distance from the galactic plane $z> 600$~pc are 
classified as thick disc stars - Fig. \ref{fig-locgal}. 

Most of our objects belong to the thin disc (over 64\%), over 
33\% stars are  thick disc stars  and a few percents of objects belong to the
galactic halo. Stars in the joint sample are generally less metal 
abundant than the Sun, with median of [Fe/H] $=-0.12$. However, the median for each 
population is different: [Fe/H] = -0.1 for thin disc, [Fe/H] = -0.16 for thick 
disc and [Fe/H] = -0.90 for halo stars.

With a substantial number of stars in both thin and thick disk the sample 
is also suitable for future analyses of planetary systems in a wider galactic perspective.

\section{Evolved stars in other planets search projects} \label{Sec_others}

\begin{figure*}
   \centering
   \includegraphics{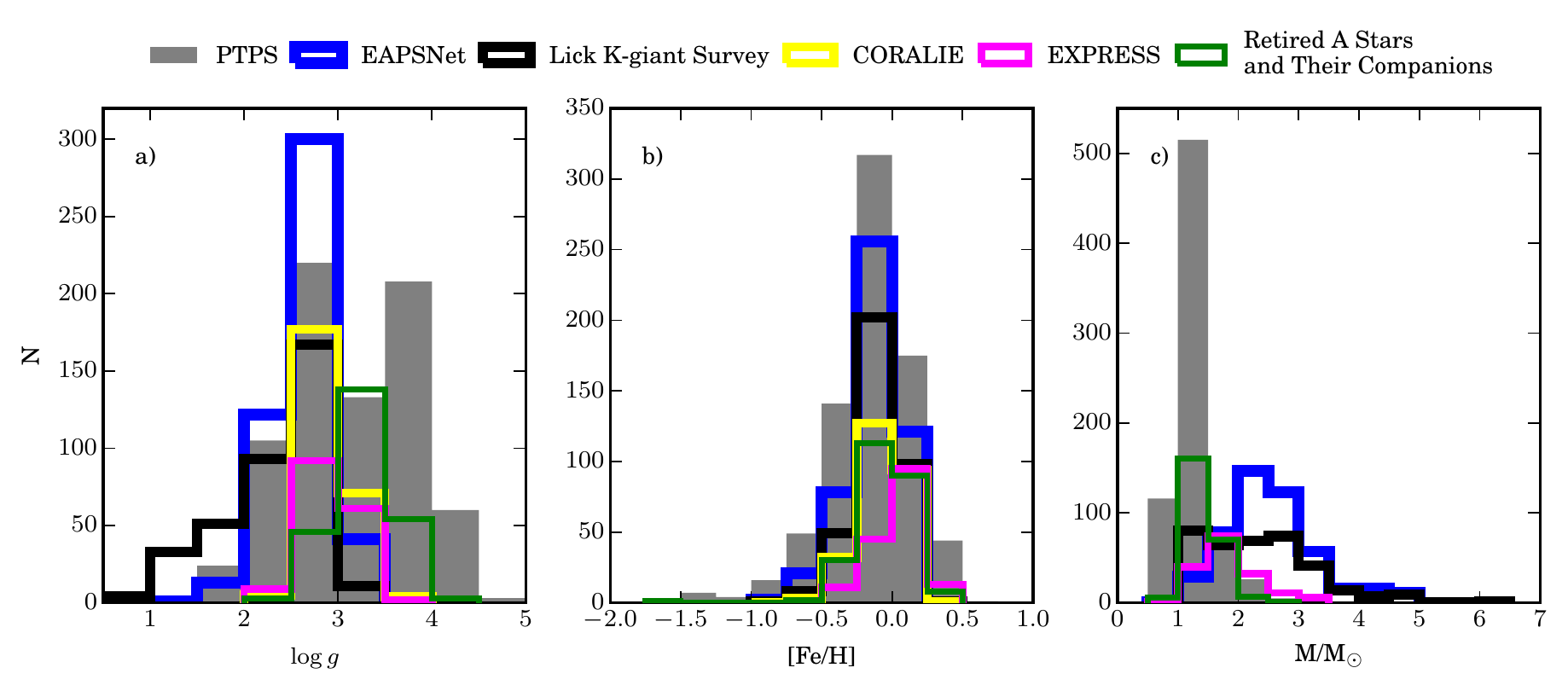}
   \caption{Histograms for $\log g$, [Fe/H] and $M/\Msun$ for planets 
                 search projects around evolved stars.}
   \label{fig-logg_Fe_M}
\end{figure*}

To our knowledge the PTPS  evolved stars sample is the largest existing sample 
of stars beyond the MS searched for low-mass companions 
with a high precision RV technique. As such it constitutes and interesting comparison 
to several other samples of stars searched for planets  around evolved stars.
In Fig.~\ref{fig-hrd-ptps} one can see that the PTPS evolved stars sample covers a wide range of 
effective temperatures, between 4000 and 6500~K, luminosity  range 
over three orders of magnitude, $\log L/L_{\odot}$ between -1 and 3, 
$\log g$ ranging from 1 to 5, and metallicity from 0.5 down to -1.5. It contains stars from both subgiant and 
red giant branches, as well a subset of highly evolved stars in horizontal branch 
or early AGB - Fig. \ref{fig-hrd-ptps}.   
The distribution of logg, [Fe/H] and masses of stars in the sample is presented in Fig. \ref{fig-logg_Fe_M}.  

To put our sample in a perspective, we
 collected available atmospheric and stellar parameters for evolved stars from several other 
planet search projects. We can this way compare our sample  with other 5 
planets search projects : 
Okayama Planet Search \citep{2003ApJ...597L.157S} and its extensions - {\it EAPSNet}, 488 stars in total for which 
stellar parameters are available in \cite{Takeda2008, Liu2010, Wang2011};
{\it Lick K-giant Survey} \citep{2002ApJ...576..478F} - stellar parameters for 359 stars presented in \cite{Reffert2014};
{\it Retired A Stars and Their Companions} \citep{Johnson2007} with stellar 
parameters for 244 stars available from \cite{Ghezzi2015};
and a radial velocity survey of 164 bright G and K giant stars in the southern hemisphere {\it EXPRESS} \citep{Jones2011},
containing altogether 1255 stars. For all these stars stellar atmospheric parameters 
as well as stellar masses, determined in various ways, are available in respective papers. 
We also added to the analysis the {\it Coralie} \& HARPS search \citep{2007A&A...472..657L}, 
with 257 stars for which atmospheric parameters (but no stellar masses) are available in \cite{Alv2015}.

A comparison of basic properties of these star samples is presented  in 
Fig.~\ref{fig-logg_Fe_M} from which 
one can easily note similarities and  differences between them. 
From Fig.~\ref{fig-logg_Fe_M}a it is clear that our evolved stars sample covers 
much wider range of stellar hosts than any other project, from bright giants to subgiants. Most of other 
samples cover the evolved stars range down to $\log g=1.5$ except 
the Lick K-giant Survey  \cite{Reffert2014} sample that extends to bright giants/super giants with $\log g$ as low as 0.5. 
In most samples, however, the largest number of stellar hosts present $\log g = 2.5-3.0$.  
The Retired A Stars and Their Companions sample \citep{Ghezzi2015} presents on the other hand 
a sample complementary to most other surveys, well consistent, however, with our sample.
Very similar is the $\Teff$ distribution with most stars in 4750-5000~K range,
again with the exception of \cite{Reffert2014} sample that extends to effective temperatures of 3750~K.

Another important feature of the samples compared here is very similar 
[Fe/H] distribution in all except  \cite{Jones2011}, which apparently contains super-solar 
[Fe/H]  objects preferentially (Fig.~\ref{fig-logg_Fe_M}b).

\begin{figure*}
  \centering
  \includegraphics{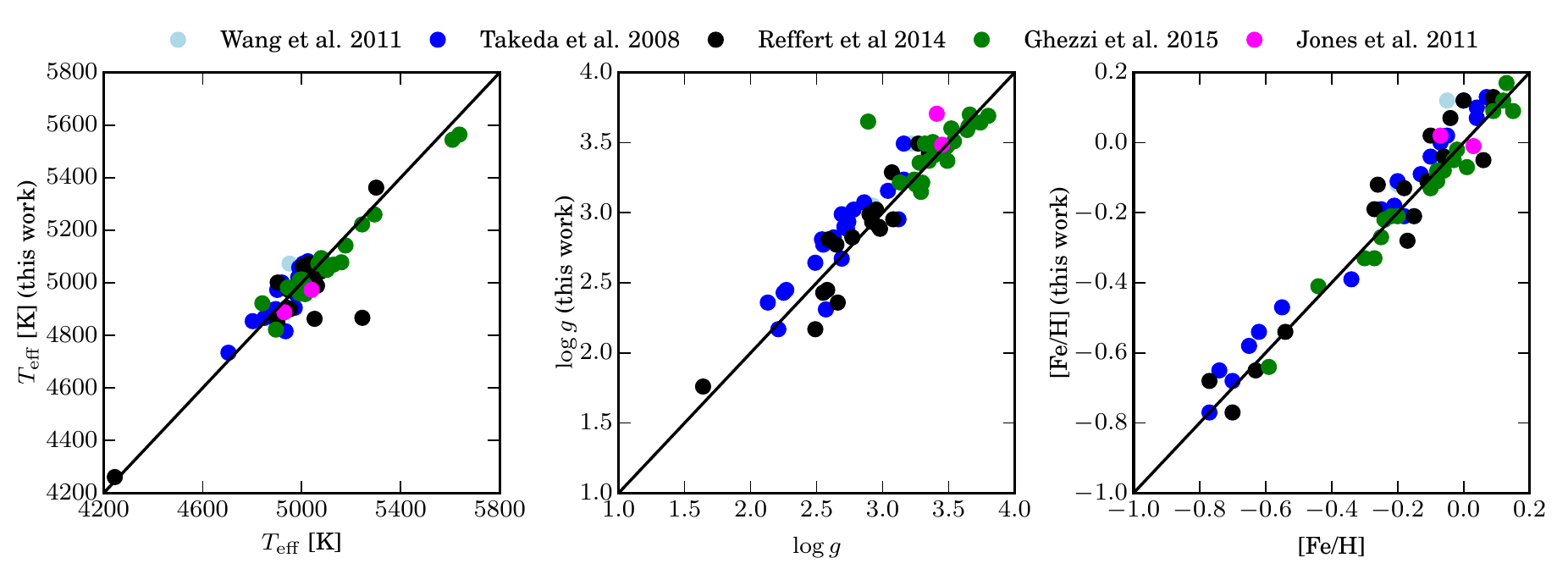}
   \caption{A comparison of stellar atmospheric for 62 PTPS stars 
               also included in other planet searches.}
   \label{fig-41common}
\end{figure*}

\begin{figure}
   \centering
   \includegraphics{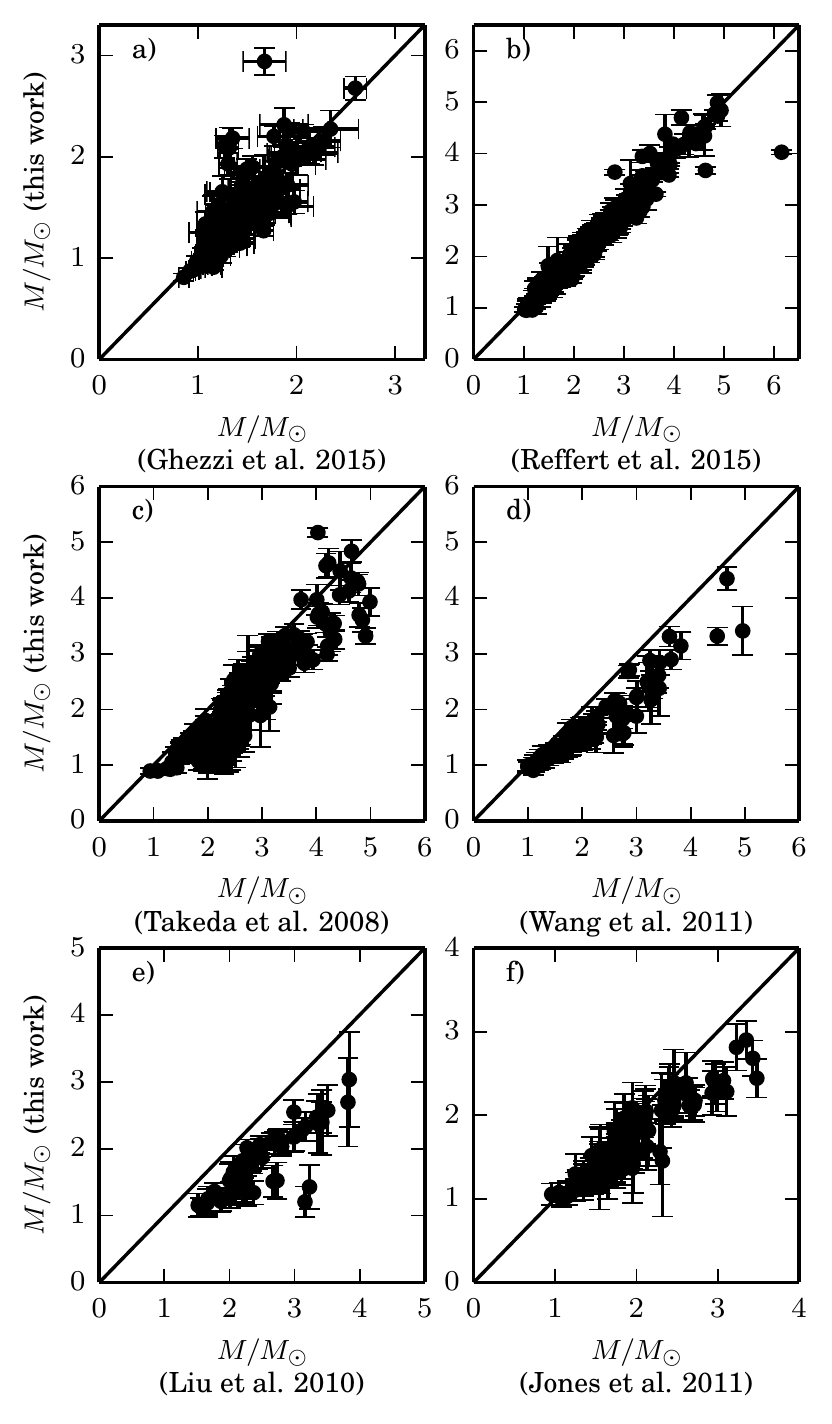}
   \caption{A comparison of stellar masses presented by respective authors 
                 with calculated by us using published stellar atmospheric parameters. 
                 Stellar masses presented by \cite{Reffert2014} Lick K-giant Survey  
                 and \cite{Ghezzi2015} Retired A Stars and Their Companions projects agree well 
                 with our estimates, while stellar masses presented by other authors do not agree well 
                 with our estimates.}
    \label{fig-Minni}
\end{figure}

In Fig. ~\ref{fig-logg_Fe_M}c we present a comparison of stellar host masses 
distributions in all six surveys. Compared to our sample, but also 
to {\it Retired A Stars and Their Companions} \citep{Ghezzi2015}  those of {\it EAPSNet} \citep{Takeda2008} 
and {\it Lick K-giant Survey} \citep{Reffert2014} contain much larger number of intermediate-mass ($M/\Msun = 2-5$) stars, 
mostly in the $M/\Msun = 2-3$ range. 
The figure suggests that the samples were defined to preferentially contain stars in that mass range. 
A quick comparison 
of mass distribution in  \cite{Takeda2008, Liu2010, Wang2011}, \cite{Reffert2014} and the presented here samples 
prove that they are especially abundant in  $2-3 M/\Msun$, $2-3 M/\Msun$ and $1-2 M/\Msun$,
respectively, making them  complementary to other planet search around evolved stars 
and potentially effective in searches for planets around stars more massive than the Sun.

In Table \ref{tab-62} we present a compilation of basic parameters for 
62 PTPS stars also included in other surveys. We note a good agreement 
in atmospheric parameters, as illustrated in Fig. \ref{fig-41common}.
However,  the agreement in stellar masses 
is not very good.

To consider in more detail stellar masses presented in various searches, we calculated masses 
for all 1255 targets from all 5 surveys with the same method we use for our data \citep{Adamczyk2015} and compared 
them to those presented in respective survey definition papers. 
Results are presented in Fig.~\ref{fig-Minni}.
We note a good agreement with \cite{Ghezzi2015} and very good with \cite{Reffert2014} results - Fig. \ref{fig-Minni}a-b. 
In the case of \cite{Reffert2014} stellar masses derived by us  
agree with their original results within $1\sigma$ on average, 
with $M_{R}/M_{our}= 1.04 \pm0.07$ (with Parsons correlation coefficient of r=0.977). 
Similar agreement between our and \cite{Ghezzi2015} data is apparent, 
with $M_{G}/M_{our}=1.05 \pm 0.29$, i.e. with noticeably larger scatter (r=0.81).

In all other cases the stellar masses derived by us are noticeably smaller.
For \cite{Takeda2008} we have $M_{T}/M_{our} = 1.29 \pm 0.24$ (r=0.911), 
for \cite{Wang2011} $M_{W}/M_{our} = 1.28 \pm0.15$ (r=0.948),
for \cite{Liu2010} $M_{L}/M_{our} = 1.43 \pm0.24$ (r=0.851), 
and for \cite{Jones2011} $M_{J}/M_{our} = 1.15\pm 0.10$ (r=0.941) (Fig~\ref{fig-Minni}c-f).
The difference between obtained here and the published stellar masses is 
in some cases quite large, on average. 

On top of that we note,  that the procedure that we use to estimate stellar masses is based 
on \cite{Bressan2012} isochrones, and assumes very moderate Reimers mass-loss with $\eta= 0.2$, 
which amount is added to the resulting final stellar mass.  In this approach the resulting stellar 
masses are MS masses actually, but as the estimated effect of mass-loss is low \citep{Miglio2012}, 
below estimated uncertainties, the simplification is justified. 
In any case our stellar mass estimated are already most likely upper limits what makes 
stellar mass estimates of the several other samples 
of \cite{Takeda2008, Liu2010, Jones2011, Wang2011}  likely to be overestimated.

\section{Discussion. Do evolved stars host more massive planets?}\label{discussion}

Using the sample presented  here and available data on other samples of evolved stars 
searched for planets we are in position to discuss one of the most intriguing features 
of the currently available sample of exoplanets around stars past the MS, 
the suspicious growth of planetary masses with evolutionary stage of theirs host \citep{Niedzielski2015b}.

Planetary systems around evolved stars are often mixed with planetary systems 
around stars more massive than the Sun. Although generally not true this is somewhat 
justified with  increase \citep{Niedzielski2015b} of average host mass 
as moving from dwarfs ($M/\Msun=0.997\pm0.016$), 
through subgiants ($M/\Msun=1.446\pm0.031$), 
to giants  ($M/\Msun=1.885\pm0.091$), or bright giants ($M/\Msun=1.464\pm0.12$).
Stellar mass increase for evolved stars with planets is not a physical phenomenon 
and it only reflects selection effects caused by the most important scientific driver 
for planets searches beyond the MS - search for planets around stars more 
massive than the Sun. A real effect, postulated by \cite{2007A&A...472..657L} 
and supported by theory by \cite{Bowler2010} might be an increase 
of planetary system mass with a stellar mass. That is, however more
 difficult to prove \citep{Niedzielski2015b}, to much extend due to uncertain masses of hosts.

The question of reliability of masses of evolved stars hosting planetary system 
was raised by  \cite{Lloyd2011, Lloyd2013, SchlaufmanWinn2013} 
and resulted in further studies  \citep{Sousa2015, Adamczyk2015}.
Meanwhile \cite{2013A&A...555A..87M} and \cite{Niedzielski2013} 
noticed that the frequency of brown dwarf (BD) companions to evolved stars seems 
to be surprisingly large, suggesting no BD desert around these stars. Indeed, \cite{Niedzielski2015b} 
showed that the average mass of a companion (according to exoplanet.eu) increases  
(RV detected companions only) from $2.25 \pm 0.1 \MJ$ around dwarfs, 
through $2.956\pm0.467 \MJ$ around subgiants, up to $6.533 \pm 1.00 \MJ$  
around giants and $7.798 \pm 1.186 \MJ$  around bright giants. 
Such an over 3 times increase in companions mass in turn of stellar evolution 
is unlikely real \citep{1998Icar..134..303D}.  It is also much faster than 
planetary system hosts average mass increase in these types of stars.

What is therefore the reason for such a fast companions mass increase in evolved stars? 
We have shown already in Sect. \ref{Sec_others} that in some planet searches around 
evolved stars stellar masses may be overestimated by up to 43$\%$. 

In Table  \ref{table-Reffert}  we present 
a compilation of stellar masses for evolved planetary system hosts from the list 
of \cite{Reffert2014} as well as all hosts of planetary systems detected within PTPS. 
The stellar masses estimated here are larger for Retired Stars and Their Companions
 and the Lick K-giants Survey. The original to estimated here stellar mass ratios are 
 $0.91 \pm0.09$ and $0.93\pm0.15$. In all other surveys stellar masses are overestimated 
 relative to our determinations and the respective mass ratios are $1.13\pm0.22$ for 
 EXPRESS, $1.42\pm0.31$ for EAPSNet and $1.14\pm0.45$ for PTPS.
On average, however, stellar masses for evolved planetary systems included 
in the compilation of \cite{Reffert2014} are only very slightly overestimated with 
original to presented here average mass ratio of $1.12\pm0.35$.

\begin{figure}
   \centering
   \includegraphics{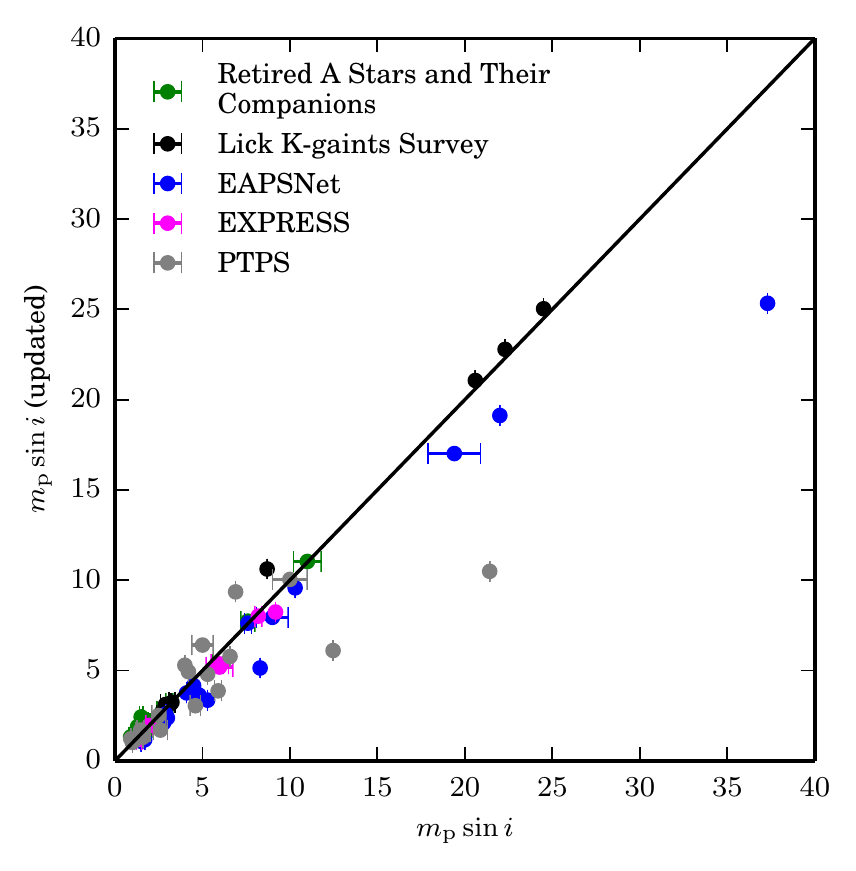}
   \caption{ A comparison of companions masses originally published and calculated
                 with our Bayesian approach. Systems 
                 listed in compilation of \cite{Reffert2014} are presented. Most companions,
                 show masses very similar to originally published. }
   \label{fig-planets_common}
\end{figure}

In Fig. \ref{fig-planets_common} we see that companions masses, recalculated with our estimates 
of hosts masses for all systems from the compilation of \cite{Reffert2014} generally agree with 
original published values, with very few exceptions. This is not surprising as the 12$\%$ overestimate 
of stellar hosts masses contributes as $\sim m^{2/3}$ to companions masses.
Therefore the apparent overestimate of stellar hosts masses in several planet search
projects does not result in overestimate of companions masses.

In Table \ref{table:new-masses} we collected from exoplanet.eu 
basic data concerning planetary systems around evolved stars ($\log g <3$) discovered 
with the RV technique according the same selection criteria as in \cite{Niedzielski2015b}. 
For all these systems we calculated 
new stellar masses of stellar host, including our own detections,  based on available 
atmospheric parameters. We also scaled masses of  their companions assuming  
the new hosts masses. We found that for individual system both increase and decrease 
of companion's masses may occur. We also calculated median, mean and standard 
deviations within $\log g=2.5\pm0.5$ (giants, 44 stars) and $\log g=1.5\pm 0.5$  (
bright giants, 16 stars) ranges. 
For giants and bright giants respectively mean, median and standard deviations 
of companions masses are 5.46, 3.33 0.84 and 7.89, 6.38 and 1.12.
We found that the low-mass companions masses around giants show lower mean masses,
compared to \cite{Niedzielski2015b}, but both values agree within $1\sigma$.
For bright giants the mean companions mass remains essentially the same. 
Again, the evident overestimate of stellar hosts masses in several planet search projects does 
not seem to be the reason of the apparent companions mass increase for evolved stars.

We conclude therefore that even with stellar masses corrected for possible overestimates 
existing data show virtually the same companions mass increase for evolved stars as presented in \cite{Niedzielski2015b}. 
We are therefore confident that the apparent companions mass increase 
for evolved stars is either real or caused by other factors than  hosts masses  overestimates.

Obviously sample definitions in planet search projects devoted to evolved stars, especially 
the abundance of relatively massive targets may  play a role.
There may be, however, more simple reasons. First of all of instrumental nature, like limited actual precision of RV measurements 
as not all surveys delivered companions in $1\MJ$ mass range.
Certainly  the stellar jitter, unresolved in time 
p-mode oscillations \citep{KjeldsenBedding1995} that in evolved stars are expected to reach $\sim 100 \ms$ 
might be another factor limiting the actual, effective RV precision when applied to luminous stars.
Although it has been demonstrated by various authors,  that with systematic observations
of such stars companions of $\leq1\MJ$ can be undoubtfully detected \citep{Gettel2012b, Gettel2012a} 
it is tempting to try to study those two factors in more detail.  

\begin{figure}
   \centering
   \includegraphics{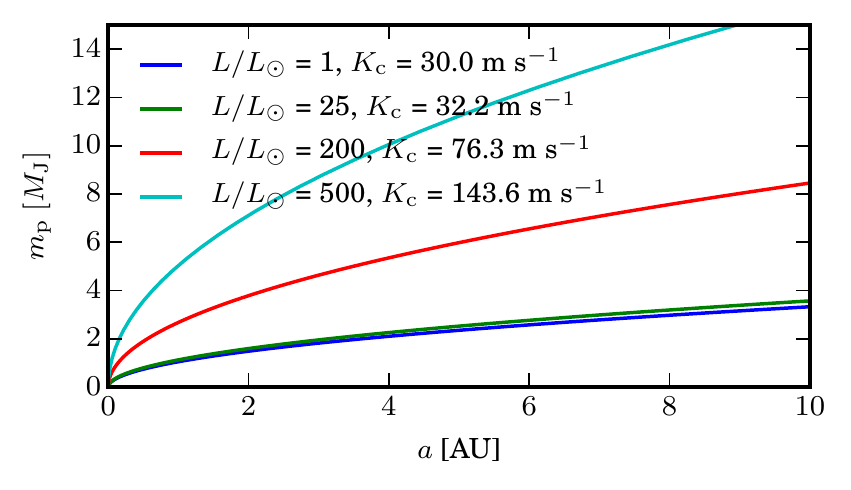}
   \caption{Minimum detectable companion masses vs. semi major axis for stars
                 at various evolutionary stages.  See text for explanation.}
   \label{fig-K}
\end{figure}

In a search for simple explanation  of the apparent low-mass companions mass increase 
for evolved stars we assumed that the effective RV precision of a planet search is:\\
 $\sigma_{e} = \sqrt{\sigma_{\mathrm{RV}}^{2}+jitter^{2}}$,\\ where $\sigma_{\mathrm{RV}}$ 
 is the instrumental RV precision and $jitter$ is the amplitude of  p-mode oscillations for given 
 stellar mass and luminosity according to \cite{KjeldsenBedding1995}. We also assumed that 
 within a planet search a low mass companion can be detected if RV semiamplitude $K \geq K_{c} = 3 \times \sigma_{e}$.
In Figure \ref{fig-K} we present the resulting $K_{c}$ at various orbital separations 
for $\sigma_{\mathrm{RV}}=10 \ms$, average stellar hosts masses for dwarfs, subgiants, giants and bright giants from \cite{Niedzielski2015b};  
and assuming (arbitrary) luminosities of  $L/L_{\odot}=1$, 25, 200 and 500, respectively.

There are two immediate conclusions obvious from Figure \ref{fig-K}: (i) the minimum detectable companion 
mass increases with orbital separation and for evolved, more active stars; and (ii) for evolved stars detection
 of nearby low-mass companions ($m_{\mathrm{p}} \leq 2 \MJ$ within 1 AU) is difficult, if at all possible.

Figure \ref{fig-K} shows that the planetary mass distributions for dwarfs and subgiants are very similar 
as the detectabillity of companions to those stars is limited by instrumental RV precision mainly. 
Moreover the  increase of minimum detectable companions masses for evolved stars comes naturally 
from the increased level of additional stellar noise.
Assuming further for simplicity a uniform distribution of companions orbit with 5 AU we obtained 
average minimum masses (in $\MJ$) of detectable companions to dwarfs, subgiants, giants and bright giants  
1.2, 1.3, 3.0 and 5.6, respectively, in qualitative agreement with  observations.

The conclusion (ii) above suggests that the apparent lack of planets within $\sim0.5$~AU around evolved stars \citep{Johnson2007, Sato2008} 
may not be a physical phenomenon but only a reflection of limited effective RV precision, if for some reason,
companions more massive than $\sim 2 \MJ$ are not so abundant in close-in  orbits around evolved stars
 (an assumption very well supported by observations). This finding seems to be also supported with recent discovery of Kepler 91 b , 
 a $0.66\pm0.06 \MJ$ planet in only $a/R_{\star}=2.253 \pm 0.046$ ($R_{\star}=6.30 \pm0.16 \Rsun$) orbit around 
 a $1.3\pm0.1 \Msun$, $\log g=2.953\pm0.07$ giant \citep{Lillo-Box2014, Barclay2015, Sato2015}.  

If  we anyway assume that low mass companions around giants and bright giants are present outside 
0.5~AU. only the average minimum companions masses to those stars  grow to 3.9 and 7.4, 
respectively, which makes our conclusion even stronger. 

Based on  instrumental precision and stellar activity only  we can at least qualitatively explain the observed 
apparent increase of companions masses for evolved stars and question the reality of the apparent lack 
of low mass companions to evolved stars within $\sim0.5$~AU.

\section{Conlusions}\label{conclusions}

We have presented atmospheric parameters, luminosities, masses, stellar ages 
and radii for 402 stars from the PTPS . We also presented estimated atmospheric 
parameters for another 53 stars, originally included in PTPS but rejected as unsuitable 
for a planet search. For 272 stars presented results are first determinations. 
We  presented and discussed in more detail the complete sample of 744 stars 
beyond the MS that form the evolved stars sample of PTPS

In a search for reason of apparent planetary mass companion increase for evolved 
stars we have compiled atmospheric parameters for 1255  stars from another 5 planet searches 
and estimated stellar masses in a uniform way. We found very good agreement 
with two Lick K-giant Survey \cite{Reffert2014} and Retired A stars and their Companions \citep{Ghezzi2015} 
but we also found that stellar masses presented in two other projects  appear to be seriously overestimated.

We do not claim that our estimates of stellar masses are better, more precise or more reliable. 
The only point we make is that the problem of stellar masses of evolved planetary systems 
hosts indeed exists. 
We stress that stellar mass estimated by us are already most likely upper limits 
what makes stellar mass estimates of the several other samples 
worth more detailed analysis.

With uniformly determined stellar masses we checked reality of the apparent planetary mass increase 
for evolved stars and we found that it is not a result of stellar mass overestimate by some authors.

We demonstrated that the apparent increase of evolved stars companion masses and the lack of planetary mass companions
 to evolved stars within 0.5 AU may both originate from limited RV precision and additional noise introduced by stellar p-mode oscillations.

\begin{acknowledgements}
We thank Dr Yoichi Takeda 
for making 
his codes available for us and Dr Luan Ghezzi for sharing his results prior to publication.
We thank the HET resident astronomers and 
telescope operators for their continuous support. 
We thank the anonymous referee for comments that let us improve the paper substantially.
MiA, AN, BS-D, MA and GN 
were supported by the Polish National Science Centre
grant no. UMO-2012/07/B/ST9/04415.
MA acknowledges the Mobility+III fellowship from the Polish Ministry of Science
and Higher Education. 
The Hobby Eberly Telescope (HET) is a joint project of the University of Texas 
at Austin, the Pennsylvania State University, Stanford University, 
Ludwig-Maximilians-Universit\"at M\"unchen, and Georg-August-Universit\"at 
G\"ottingen. The HET is named in honor of its principal benefactors, William P. 
Hobby and Robert E. Eberly. The Center for Exoplanets and Habitable Worlds is 
supported by the Pennsylvania State University, the Eberly College of Science, 
and the Pennsylvania Space Grant Consortium. This research has made extensive 
use of the SIMBAD database, operated at CDS (Strasbourg, France) and NASA's 
Astrophysics Data System Bibliographic Services.
\end{acknowledgements}

\bibliographystyle{aa} 
\bibliography{literature} 

\longtab{1}{

}

\longtab{1}{
\begin{landscape}
\scriptsize{
\begin{longtable}{l@{\hspace{1.6mm}}l@{\hspace{1.6mm}}r@{\hspace{1.6mm}}r@{\hspace{1.6mm}}r@{\hspace{1.6mm}}r@{\hspace{1.6mm}}r@{\hspace{1.6mm}}r@{\hspace{1.6mm}}r@{\hspace{1.6mm}}r@{\hspace{1.6mm}}r@{\hspace{1.6mm}}r@{\hspace{1.6mm}}r@{\hspace{1.6mm}}r@{\hspace{1.6mm}}r@{\hspace{1.6mm}}r@{\hspace{1.6mm}}r@{\hspace{1.6mm}}r@{\hspace{1.6mm}}r}
\caption{\label{tab-res} Derived physical parameters for the PTPS Red  Giants Clump sample stars.}\\
\hline\hline
Ident & Ident & V	& $\rm{B-V}$  & $\pi$ & $\Teff$   &    $\log g$  &    $\vt$  &    [Fe/H]  & $v\sin i_*$ & $RV$  &   $MJD$  &  $(\rm{B-V})_{0}$   &  $M_{\rm{V}}$   &   $BC_{\rm{V}}$   &  $\log L/L_{\odot}$  &    $M$ &    $R$ &  $\log \mathrm{age}$ \\ 
 TYC &    HD or BD & [mag]  &  [mag]   &   [mas] &  [K]  &    &  [$\kms$]  &   & [$\kms$] & [$\kms$]  &  +2400000 [d] &    [mag]  & [mag]  &  [mag]   &   &  [$\Msun$]  & [$\Rsun$] & [yr] \\
\hline
\endfirsthead
\caption{continued.}\\
\hline\hline
Ident  & Ident & V	& $\rm{B-V}$  & $\pi$ &  $\Teff$   &    $\log g$  &    $\vt$  &    [Fe/H]  & $v\sin i_*$ & $RV$  &   $MJD$  &  $(\rm{B-V})_{0}$   &  $M_{\rm{V}}$   &   $BC_{\rm{V}}$   &  $\log L/L_{\odot}$  &    $M$ &    $R$ &  $\log \mathrm{age}$ \\ 
 TYC &  HD or BD &   [mag]  &  [mag]   &   [mas] &  [K]  &    &  [$\kms$] &   &  [$\kms$] & [$\kms$]  &  +2400000 [d] &    [mag]  & [mag]  &  [mag]   &   &  [$\Msun$]  & [$\Rsun$] & [yr] \\
\hline
\endhead
\hline
\endfoot
0003-01118-1 & HD 2221 & 8.42 & 0.780 $\pm$ 0.030 & 10.60 $\pm$ 0.93 & 5291 $\pm$ 5 & 3.77 $\pm$ 0.01 & 0.93 $\pm$ 0.05 & -0.13 $\pm$ 0.01 & 1.19 $\pm$ 0.39 & -19.603 $\pm$ 0.025 & 54428.10088 & 0.761 & 3.48 & -0.181 & 0.58 $\pm$ 0.13 & 1.07 $\pm$ 0.04 & 2.28 $\pm$ 0.26 & 9.85 $\pm$ 0.06 \\
0011-00529-1 & HD 4352 & 7.68 & 0.940 $\pm$ 0.030 & 6.57 $\pm$ 0.70 & 4878 $\pm$ 5 & 3.00 $\pm$ 0.01 & 1.09 $\pm$ 0.04 & -0.28 $\pm$ 0.01 & 1.54 $\pm$ 0.48 & -11.644 $\pm$ 0.024 & 53985.32437 & 0.930 & 1.69 & -0.305 & 1.35 $\pm$ 0.15 & 1.01 $\pm$ 0.04 & 5.94 $\pm$ 0.74 & 9.94 $\pm$ 0.07 \\
0019-01406-1 & HD 6734 & 6.46 & 0.850 $\pm$ 0.020 & 20.63 $\pm$ 0.45 & 4986 $\pm$ 15 & 3.54 $\pm$ 0.04 & 0.85 $\pm$ 0.06 & -0.47 $\pm$ 0.01 & 1.86 $\pm$ 1.27 & -94.760 $\pm$ 0.027 & 54000.39176 & 0.851 & 3.03 & -0.269 & 0.79 $\pm$ 0.07 & 0.90 $\pm$ 0.03 & 2.99 $\pm$ 0.37 & 10.07 $\pm$ 0.05 \\
0048-00271-1 & HD 18012 & 6.75 & 0.900 $\pm$ 0.020 & 8.14 $\pm$ 0.80 & 4790 $\pm$ 5 & 2.68 $\pm$ 0.03 & 1.12 $\pm$ 0.04 & -0.67 $\pm$ 0.01 & 1.02 $\pm$ 0.39 & -124.812 $\pm$ 0.027 & 53757.11663 & 0.926 & 1.30 & -0.342 & 1.52 $\pm$ 0.13 & 0.88 $\pm$ 0.03 & 7.73 $\pm$ 1.08 & 10.07 $\pm$ 0.05 \\
0052-00372-1 & HD 16466 & 8.41 & 0.805 $\pm$ 0.037 & 12.16 $\pm$ 1.31 & 5474 $\pm$ 12 & 4.23 $\pm$ 0.04 & 0.90 $\pm$ 0.08 & 0.02 $\pm$ 0.02 & 1.68 $\pm$ 0.64 & 47.583 $\pm$ 0.026 & 53996.31227 & 0.712 & 3.76 & -0.275 & 0.50 $\pm$ 0.24 & 0.94 $\pm$ 0.02 & 1.60 $\pm$ 0.37 & 10.10 $\pm$ 0.03 \\
0061-01209-1 & HD 18747 & 8.13 & 0.940 $\pm$ 0.041 & 12.99 $\pm$ 1.40 & 4994 $\pm$ 10 & 3.69 $\pm$ 0.04 & 0.86 $\pm$ 0.06 & -0.03 $\pm$ 0.01 & 0.0 $\pm$ 0.51 & 47.501 $\pm$ 0.041 & 54081.25039 & 0.904 & 3.62 & -0.263 & 0.56 $\pm$ 0.17 & 1.10 $\pm$ 0.07 & 2.52 $\pm$ 0.48 & 9.88 $\pm$ 0.10 \\
\end{longtable}
}
\end{landscape}
}

\begin{table*}
\caption{An photometric estimate of $\Teff$ and $\log g$  for stars with unstable CCF.}
\label{table-rejected}

\centering
\begin{tabular} {lllll}
\hline\hline
Name & HD/BD   & V [mag] & Teff [K] & logg \\ \hline
\multicolumn{5}{c}{Spectroscopic binaries - SB2}   \\ \hline
TYC~0738-01228-1 & HD 42071 & 8.36 & 5828 $\pm$ 23 & 4.40 $\pm$ 0.54 \\
TYC~0835-00046-1 & BD+14 2168 & 9.37 & 4709 $\pm$ 38 & 2.84 $\pm$ 0.51 \\
TYC~0928-00468-1 & HD 138750 & 7.88 & 6032 $\pm$ 21 & 4.37 $\pm$ 1.06 \\
TYC~1142-00238-1 & HD 212141 & 8.22 & 5695 $\pm$ 21 & 4.43 $\pm$ 0.70 \\
TYC~1181-01709-1 & HD 483 & 6.99 & 5838 $\pm$ 21 & 4.40 $\pm$ 0.57 \\
TYC~1188-01246-1 & HD 4077 & 8.15 & 6231 $\pm$ 22 & 4.33 $\pm$ 1.11 \\
\hline
\end{tabular}
\end{table*}

\longtab{3}{
\scriptsize{
\begin{longtable}{lllllllllll}
\caption{\label{tab-62} 62 PTPS stars also included in other surveys.}\\
\hline\hline
Name           & $\Teff$ [K]  & $\log g$ & [Fe/H] & $M/\Msun$ & $\log L/L_{\odot}$ &  $\Teff$ [K]  & $\log g$ & [Fe/H] & $M/\Msun$ & $\log L/L_{\odot}$ \\
\hline
\hline
               &  \multicolumn{5}{c}{ Jones et al. (2011)}                      & \multicolumn{5}{c}{ this work}\\
 \hline
TYC~4717-01094-1  &  4930  &  3.41  &  -0.07 $\pm$ 0.07 &  1.05  &  0.62  &  4888   $\pm$ 10  &  3.71 $\pm$ 0.04  &  0.02 $\pm$ 0.02 &  0.995 $\pm$ 0.03 &  0.60  $\pm$ 0.07\\
TYC~4989-00183-1  &  5040  &  3.45  &  0.03 $\pm$ 0.12 &  1.51  &  0.86  &  4973  $\pm$ 12 &  3.49 $\pm$ 0.04  &  -0.01 $\pm$  0.02 &  1.27 $\pm$  0.06 &  0.85 $\pm$ 0.12 \\
\hline
               &  \multicolumn{5}{c}{ Wang et al. (2011)}                      & \multicolumn{5}{c}{ this work}\\
\hline
TYC~1488-00278-1  &  4933  &  2.93  &  -0.20  &  1.77  &  1.47  &  4977 $\pm$ 15 &  3.05 $\pm$ 0.05  &  -0.12 $\pm$ 0.02 &  1.57 $\pm$ 0.10  &  1.42 $\pm$ 0.08 \\
TYC~3147-01921-1  &  4951  &  3.22  &  -0.05  &  1.65  &  1.23  &  5073  $\pm$ 17 &  3.49 $\pm$ 0.05   &  0.12 $\pm$ 0.02  &  1.74 $\pm$ 0.05  &  1.14 $\pm$ 0.06 \\
\hline
\end{longtable}
}

}

\begin{table*}
\caption{ A comparison of stellar hosts masses for planetary systems from the list of \cite{Reffert2014} originally published by various authors with our estimates.}

\label{table-Reffert}

\centering

\begin{tabular} {lll}

\hline\hline

Name & Mass (original paper) & Mass (this work) \\
 & \multicolumn{2}{c} {[$\Msun$]}   \\ 
 \hline
 \multicolumn{3}{c} {EXPRESS}  \\ \hline
HIP 63242  &  1.54  $\pm$  0.05   &  1.31  $\pm$  0.20  \\ 
HIP 105854  &  2.12  $\pm$  0.10   &  2.05  $\pm$  0.15  \\ 
HIP 67851   &  1.63  $\pm$  0.20   &  1.32  $\pm$  0.16  \\ 
HIP 65891  &  2.50  $\pm$  0.21   &  2.11  $\pm$  0.14  \\ 
HIP 107773  &  2.42  $\pm$  0.27   &  2.39  $\pm$  0.40  \\
\hline

\hline

\end{tabular}
\end{table*}

\begin{table*}
\caption{\label{table:new-masses}  Original and updated stellar atmospheric parameters for 57 RV detected evolved stars ($\log g<3$), including 11 objects from red clump PTPS sample stars.}
\begin{tabular}{lcccccccc}
\hline
Planet & [Fe/H] & $\Teff$~[K] & $\log g$ & $\log L/L_{\odot}$ & $m_{\rm{p}} \sin i$ & $M_\star$ & $M_\star$(updated) & $m_{\rm{p}} \sin i$ (updated) \\
\hline
11 UMi b & 0.04 $\pm$ 0.04 & 4340 $\pm$ 70 & 1.60 $\pm$ 0.15 & 2.36 $\pm$ 0.08 & 10.50 & 1.80 & 1.74 $\pm$ 0.23 & 10.27 \\
42 Dra b & -0.46 $\pm$ 0.05 & 4200 $\pm$ 70 & 1.71 $\pm$ 0.05 & 2.26 $\pm$ 0.08 & 3.88 & 0.98 & 1.01 $\pm$ 0.09 & 3.95 \\
Aldebaran b & -0.27 $\pm$ 0.05 & 4055 $\pm$ 70 & 1.20 $\pm$ 0.10 & 2.58 $\pm$ 0.08 & 6.47 & 1.13 & 1.10 $\pm$ 0.13 & 6.37 \\
HD 122430 b & -0.05 $\pm$ 0.05 & 4300 $\pm$ 70 & 2.00 $\pm$ 0.20 & 2.27 $\pm$ 0.10 & 3.71 & 1.39 & 1.59 $\pm$ 0.22 & 4.06 \\
HD 208527 b & -0.09 $\pm$ 0.16 & 4035 $\pm$ 65 & 1.60 $\pm$ 0.30 & 2.98 $\pm$ 0.21 & 9.90 & 1.60 & 2.15 $\pm$ 0.78 & 12.07 \\
HD 220074 b & -0.25 $\pm$ 0.25 & 3935 $\pm$ 110 & 1.30 $\pm$ 0.50 & 2.84 $\pm$ 0.20 & 11.10 & 1.20 & 1.40 $\pm$ 0.31 & 12.29 \\
HD 47536 b$^{a)}$ & -0.68 $\pm$ 0.05 & 4352 $\pm$ 70 & 2.10 $\pm$ 0.20 & 2.25 $\pm$ 0.09 & 4.96 & 1.10 & 1.01 $\pm$ 0.11 & 4.69 \\
\end{tabular}
\end{table*}

\end{document}